\documentclass[draft]{agujournal2019}
\usepackage{apacite}
\usepackage{url} 
\usepackage{lineno} 
\usepackage{soul}

\draftfalse

\journalname{JGR-Space Physics}

\begin{document}

%
%

\title{Termination of Solar Cycles and Correlated Tropospheric Variability}

%
%




\authors{Robert J. Leamon\affil{1,2},
Scott W. McIntosh\affil{3}, and
Daniel R. Marsh\affil{3,4}
}

\affiliation{1}{University of Maryland, Dept. of Astronomy, College Park, MD 20742, USA.}
\affiliation{2}{NASA Goddard Space Flight Center/ Code 672, Greenbelt, MD 20771, USA.}
\affiliation{3}{High Altitude Observatory, National Center for Atmospheric Research, P.O. Box 3000, Boulder CO, 80307, USA.}
\affiliation{4}{Atmospheric Chemistry Observation and Modeling Laboratory, National Center for Atmospheric Research, P.O. Box 3000, Boulder CO, 80307, USA.}




\correspondingauthor{Robert J.\ Leamon}{robert.j.leamon@nasa.gov}




\begin{keypoints}
\item A solar cycle's fiducial clock does not run from the canonical min or max, instead resetting when old cycle flux is gone.
\item Many cycles indicate that ENSO is correlated to changes in the cosmic ray flux over the cycle. 
\item Cycle 24 is projected to end in mid 2020. Based on historical correlations, we anticipate a persisting El Ni\~{n}o in 2019, and a strong La Ni\~{n}a in 2020--21. 
\end{keypoints}


%
%



\begin{abstract}
The Sun provides the energy required to sustain life on Earth and drive our planet's atmospheric circulation. 
However, establishing a solid physical connection between solar and tropospheric variability has posed a considerable challenge across the spectrum of Earth-system science. 
The canon of solar variability, the solar fiducial clock, lies almost exclusively with the 400 years of human telescopic observations that demonstrates the waxing and waning number of sunspots, 
over an 11(ish) year period. 
Recent research has demonstrated the critical importance of the underlying 22-year magnetic polarity cycle in establishing the shorter sunspot cycle. 
Integral to the manifestation of the latter is the spatio-temporal overlapping and migration of oppositely polarized magnetic bands. The points when these bands emerge at high solar latitudes and 
cancel
at the 
equator are separated by almost 20 years. 
Here we demonstrate the impact of these ``termination'' points on the Sun's radiative output and particulate shielding of our 
atmosphere through the dramatically rapid reconfiguration of solar magnetism. 
These events reset the Sun's fiducial clock and present a new portal to explore the Sun-Earth connection. 
Using direct observation and proxies of solar activity going back 
six decades we can, with high statistical significance, demonstrate an apparent correlation between the 
solar cycle terminations 
and the largest swings of Earth's oceanic indices---a previously overlooked correspondence. 
Forecasting
the Sun's global behavior places the next solar termination 
in early 2020; should a major oceanic swing follow, 
our
challenge becomes: when does correlation become causation and how does the process work?
\end{abstract}

%
%

\section{Introduction}

Establishing a solid, physical, link between solar and tropospheric variability across timescales has posed a considerable challenge, despite the broad acknowledgment that the Sun provides the underlying energy to drive weather and climate \cite{2010RvGeo..48.4001G}. 
As it stands, solar connections to decadal-scale massive shifts in terrestrial weather patterns, like those of the North Atlantic Oscillation 
[NAO; \citeA{1995Sci...269..676H}], 
or the El Ni\~{n}o Southern Oscillation 
[ENSO; \citeA{1997BAMS...78.2771T,2009Sci...325.1114M}] 
are little more than anecdotal. 
ENSO 
is the combination of three related climatological phenomena, El Ni\~{n}o, La Ni\~{n}a and the large-scale seesaw exchange of sea level air pressure between areas of the western and southeastern Pacific Ocean \cite{2014GeoRL..41..161V}. 

The ``normal''
circulation pattern over the tropical Pacific is known as the ``Walker Circulation,'' 
after Sir Gilbert Walker, who first described it as an employee of the British Meteorological Office in India.
The Southern Oscillation describes a bimodal variation in sea level barometric pressure between observation stations at Darwin, Australia and Tahiti.  It is quantified in the Southern Oscillation Index (SOI), which is a standardized difference between the two barometric pressures.  
Normally, lower pressure over Darwin and higher pressure over Tahiti encourages a circulation of air from east to west, drawing warm surface water westward and bringing precipitation to Australia and the western Pacific, and a return of west-to-east flow in the upper troposphere.  
When the pressure difference weakens, which is strongly coincidental with strong positive (warm) phases of ENSO, El Ni\~{n}o conditions (increased rainfall in California and the Gulf Coast states, dry Midwest and Eastern seaboard, and a very hot and dry Australia) occur; 
La Ni\~{n}a is the opposite, when the Walker circulation is strong. 
(Topically, 
the Atlantic hurricane season tends to be more active during La Ni\~{n}a years, due to reduced upper-level vertical wind shear.
Conversely, El Ni\~{n}o favors stronger hurricane activity in the central and eastern Pacific basins.)

{\color{blue}
Combining the costs of natural disaster recovery with the costs associated with yields of major commodity crops \cite{IizumiEA14,2017PLoSO..1279086G},
the need to be able to predict ENSO events beyond a seasonal 
forecast\footnote{{\em e.g.}, {\tt http://www.wmo.int/pages/prog/wcp/wcasp/enso\_update\_latest.html}} is high.
Flooding in Australia 
during the 2010--12 La Ni\~{n}as 
and the ensuing economic cleanup costs (A\$5--10 billion (US\$4.9--9.8 billion) in Queensland alone) 
led to the commissioning of a Government Report on ``two of the most significant events in Australia's recorded meteorological history'' \cite{2012AusBOM..LaNina}.
Similarly, the Peruvian government estimated the very strong 1997--1998 El Ni\~{n}o event cost about US\$3.5 billion, or about 5\% of their gross domestic product (GDP).
In the United States, the National Oceanic and Atmospheric Administration (NOAA) assessed direct economic losses from that event at US\$34 billion, with a loss of 24,000 lives.
(And US\$34 billion in 1998 becomes US\$52.5 billion in 2018 dollars assuming CPI inflation.)
That assessment comes with the important caveat that 
losses associated with El Ni\~{n}o-related floods or droughts in some areas can be offset by gains elsewhere, for instance through reduced North Atlantic hurricane activity, lower winter heating bills or better harvests for certain crops---Argentinian wheat yields are strongly increased in El Ni\~{n}o years, for example, whereas US (and moreso Canadian) yields fall \cite{2017PLoSO..1279086G}.

That crop yields in North and South America, Australia and Eurasia vary, along with regional temperature and precipitation changes, 
makes it clear that
ENSO influences, through ``teleconnections,'' 
\cite<e.g.,>{2019RvGeo..57....5D}
the global dynamics of seasonal winds, rainfall and temperature. 
These teleconnections imply, indeed require, coupling throughout the atmosphere, and despite the mention of troposphere in the title of this paper, manifestations of ENSO are observed throughout the neutral atmosphere and higher.
}

\subsection{Solar Cycle Modulations}

ENSO has been suggested to be a significant source of tidal variability in the mesophere and lower thermosphere \cite<MLT;>{2005GeoRL..3213805G,2007JGRD..11220110L}.
These tides can be modulated by tropospheric forcing.
\citeA{2005GeoRL..3213805G} suggested that the large-scale convective systems originating over the western Pacific facilitate excitation of nonmigrating tides through latent hear release or large-scale redistribution of water vapor that compete with the dominant migrating tide and possibly induce the observed interannual variability in the diurnal tide.
\cite{2007JGRD..11220110L} noted a pronounced `spike' in diurnal tide amplitude in the central Pacific in late 1997 and early 1998 and linked the phenomenon to ENSO, through water vapor absorption and diurnal latent heat release due to deep convection.
\citeA{2018EP&S...70...85S} showed that the propagating diurnal temperature tides in the MLT (altitudes of $\sim$100km) were also sensitive to ENSO effects. 
Numerous studies have also reported that ENSO controls the stratospheric quasi-biennial oscillation (QBO) through the interactions of broadband atmospheric waves and mean flows.
Although the QBO is a tropical phenomenon, it affects the stratospheric flow from pole to pole by modulating the effects of extratropical waves. 
El Ni\~{n}o events shorten the QBO period to $\sim$2 years, whereas the period lengthens to $\sim$2.5 years during La Ni\~{n}a events and neutral times.
The QBO has long been tied to the solar cycle
\cite{1987GeoRL..14..535L,1988JATP...50..197L,2005JASTP..67...45L}:
the arctic middle stratosphere tends to be colder and the polar vortex stronger when the equatorial wind is westerly than when it is easterly,  
when the QBO is in its west 
phase, 
and that atmospheric correlations with solar cycle variability are stronger if the data are sorted by east- or west-phase of the QBO.

The 11-ish year activity cycle is connected with a large variability of the solar radiation in the ultraviolet (UV) part of the spectrum which varies about 6--8\% between solar maxima and minima \cite{1994JGR....9920665C,2000GeoRL..27.2425L}. 
\citeA{2009Sci...325.1114M} 
showed that the top-down stratospheric response of ozone to UV solar forcing, when combined with bottom-up coupled ocean-atmosphere response, could amplify the small TSI fluctuations into the observed sea surface temperature variations; enhanced UV radiation stimulates additional stratospheric ozone production, and thus UV absorption, thus warming that later differentially with respect to latitude. 
That is enough to cause in the upper stratosphere changes in the temperatures, winds and ozone which will result in circulation changes here and it is possible that such changes have an indirect effect on the lower stratosphere and on the troposphere.
One bottom-up forcing mechanism is again a positive feedback loop whereby greater solar energy input falls on the relatively cloud-free tropical oceans, which then evaporate more moisture, which is then carried by trade winds to convergence zones and precipitates. This precipitation strengthens that Hadley and Walker circulations, increasing the trade winds.
In addition to a low-frequency (linear) response, the solar cycle variation of UV introduces a high-frequency (nonlinear) response that is considerably stronger.
Involving interannual fluctuations, the high-frequency response is associated
with the QBO and its influence on the Brewer-Dobson circulation \cite{2002JGRD..107.4749K,2006JGRD..111.6110S}.

\begin{figure}[tbp]
\centering
\includegraphics[width=0.4875\linewidth]{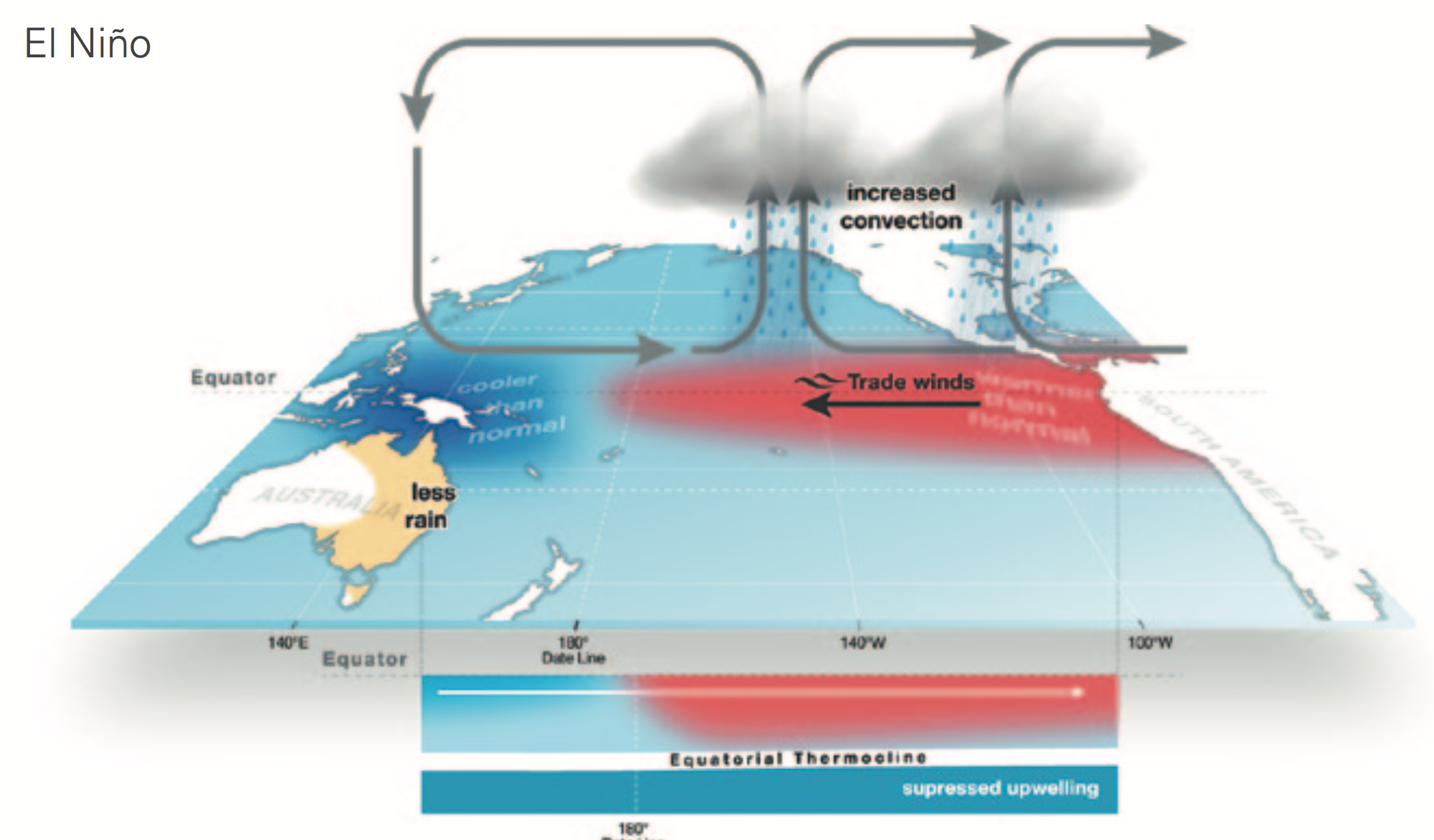}
\includegraphics[width=0.4875\linewidth]{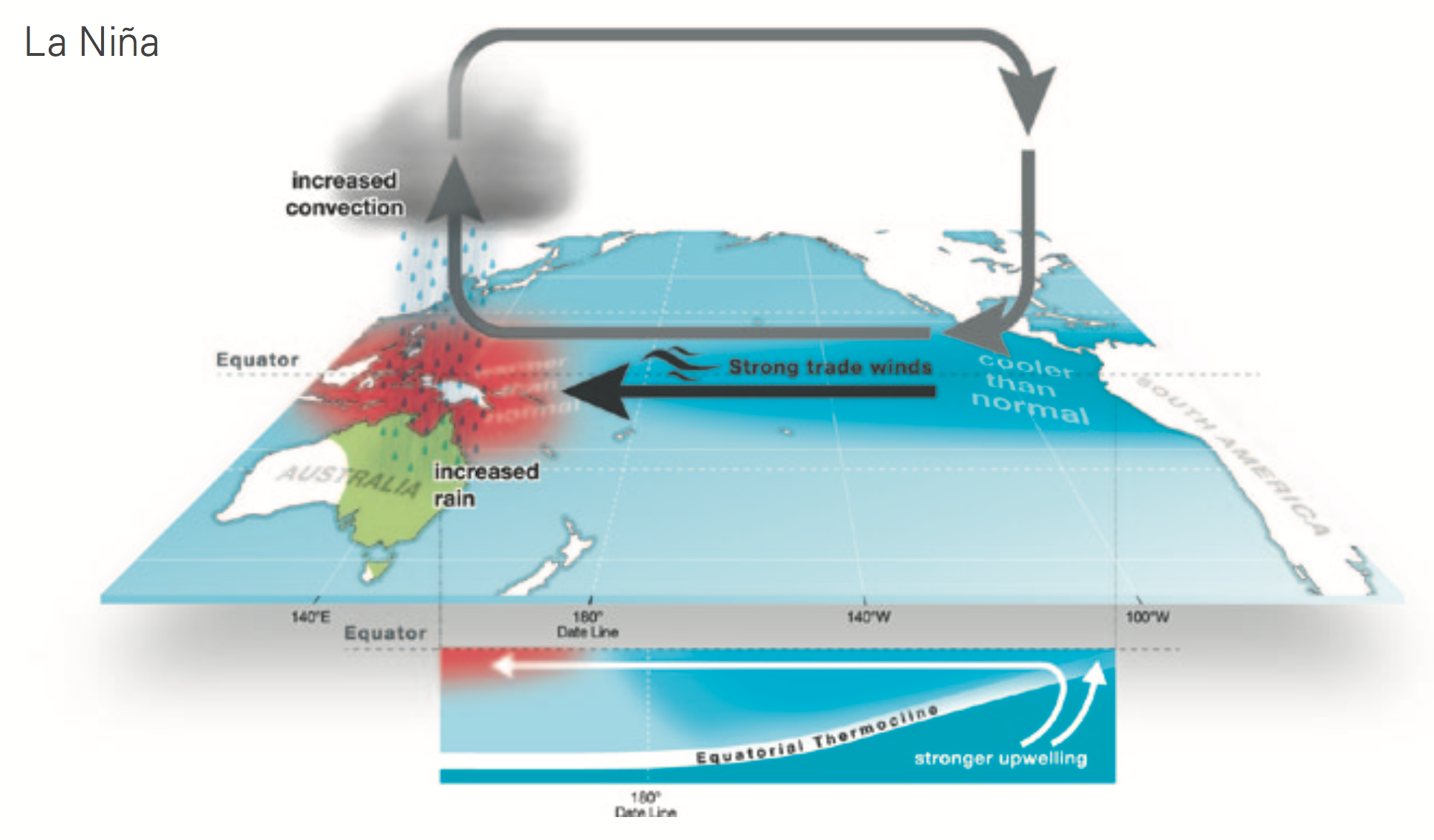}
\includegraphics[width=0.4875\linewidth]{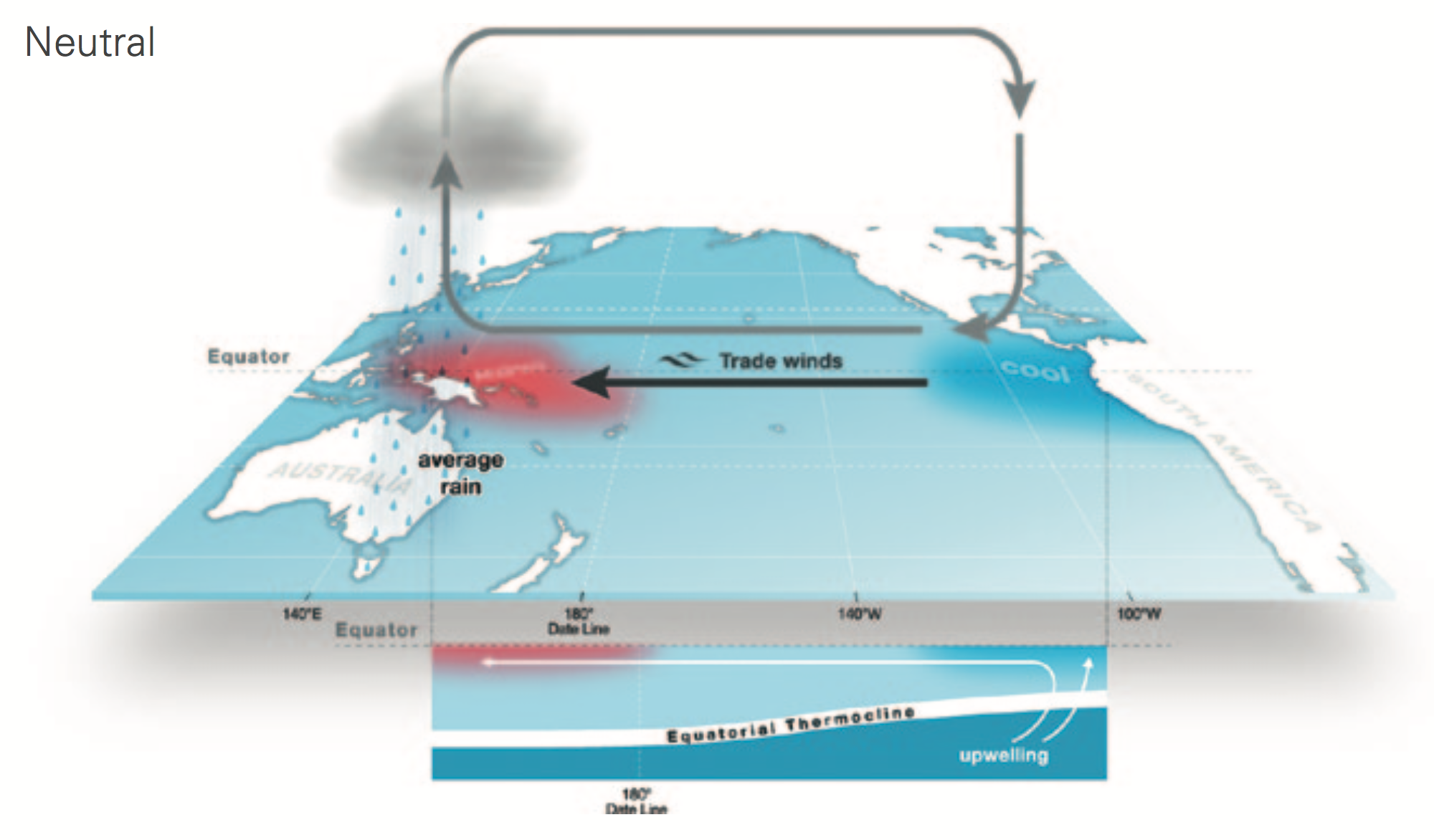}
\caption{Typical circulation patterns during El Ni\~{n}o (left), La Ni\~{n}a (right), and neutral atmosphere (below middle).
Reproduced from Australian \protect\citeA{2012AusBOM..LaNina}.
}
\label{fig:wmo}
\end{figure}

The connection between the Southern Oscillation and precipitation is also manifest in the quantity of long-wave (e.g., infrared) radiation leaving the atmosphere.  
Under clear skies, a great deal of the long-wave radiation released into the atmosphere from the surface can escape into space.  
Under cloudy skies, some of this radiation is prevented from escaping.  
Satellites are able to measure the amount of long-wave radiation reaching space, and from these observations, the relative amount of convection in different parts of the basin can be estimated.
\citeA{2017JGRC..122.7880P} discussed the impact of ENSO on surface radiative fluxes, concluding that the maximum variance of anomalous incoming solar radiation is located just west of the dateline and coincides with the area of the largest anomalous SST gradient, reaching up to 60 W/m$^2$ and lagging behind the Ni\~{n}o3 index by about a month, suggesting a response to anomalous SST gradient, 
{\em i.e.}, ENSO is not driven by changes in insolation.
{\color{blue}
(We shall return to \citeA{2017JGRC..122.7880P} and 
the implications of 
temporal correlations and causality 
below.)
}

An 11-year dependence on ENSO \cite{2009Sci...325.1114M,2004JASTP..66.1767V} 
and other atmospheric, oceanic, and climate phenomena including enhanced summer monsoon precipitation over India \cite{2004GeoRL..3124209K,2012GeoRL..3913701V} and the NAO, have been well documented in the literature. 
However, 
investigations 
linking decadal-scale tropospheric activity with those of the Sun have relied on the canon of the 11-year solar activity cycle \cite{2015LRSP...12....4H}, requiring (apparently ad-hoc) mathematical phase shifts to be introduced to establish {\em any\/} kind of link \cite{2004JASTP..66.1767V, doi:10.1029/2006JC004057, 2010RvGeo..48.4001G, doi:10.1029/2011GL047964}. 
It is fair, then, to say that searching for the connection between the variability of the solar atmosphere and that of our troposphere has become ``third-rail science''---not to be touched at any cost.
Nevertheless, 
this paper will show the strong likelihood of a solar driver of ENSO, 
and 
some measure of 
prediction skill over the coming decade ({\em i.e.}, solar cycle 25). 

\subsection{Solar Cycle Terminators}

Recent studies highlighting the presence, and traceability, of the twenty-two year magnetic cycle of the Sun have revealed the occurrence a new type of event in the solar lexicon---the ``Terminator'' \cite{M19,2019NatSR...9.2035D,1984Terminator} 
Stated simply, a terminator is the event that marks the hand-over from one sunspot cycle to the next. It is an abrupt event occurring at the solar equator resulting from the annihilation/cancellation of the oppositely polarized magnetic activity bands at the heart of the 22-year cycle; i.e., there is no more old cycle flux left on the disk. 
This annihilation appears to globally modify the conditions for magnetic flux to emerge---principally causing the rapid growth of the magnetic system at mid solar latitudes that will be the host for the sunspots of the next sunspot cycle. 
{
Put another way, rather than thinking of a solar cycle beginning or ending at the minimum of the sunspot number record\footnote{And the 13-month smoothed SSN record, so by rigorous definition the minimum cannot be defined until a year after it has occurred.}, the terminators define the end of influence of the old cycle on the Sun and solar output.
}
Our companion paper \cite[{hereafter M2019}]{M19}
highlights the terminators that took place at the end of solar cycles~22 and~23, illustrating that a significant, step-function-like, change in the Sun's radiative proxies took place at the same time over a matter of only a few days. In their analysis, M2019 demonstrate that terminators were visible in standard proxies of solar activity going back many decades---as many as 140 years to the dawn of synoptic  H-$\alpha$ filament and sunspot observations. 
\citeA{2019NatSR...9.2035D} suggested that the most plausible
mechanism for rapid transport of information 
from the equatorial termination of the old cycle's activity bands 
(of opposite polarity in opposite hemispheres)
to the mid-latitudes to trigger new-cycle growth
was a solar ``tsunami'' in the solar tachocline that migrates poleward with a gravity wave speed ($\sim$300$\mbox{ km s}^{-1}$).

In the following analysis we will explore {\em if\/} the signature of these solar termination events could provide a starting point in establishing a robust Sun-Troposphere connection on decadal timescales---by creating a new fiducial time for solar activity. 
We start with episodes of largest fluctuation in the El Ni\~{n}o Southern Oscillation, the so-called El Ni\~{n}o ``events.'' Methodically assessing the solar observations we will  highlight the radiative and particulate signatures of the two best sampled termination events---the two most recent in 1997 and 2010-11. Using standard measures of solar variability over decades we can extend to the dawn of the space age---where the proxy data is most reliable. Following the introduction of a data-inspired schematic view of the Sun's 22-year magnetic activity cycle over that same we will draw comparison with the ocean index. Employing a modified version of the Superposed Epoch Analysis 
[SEA; \citeA{1913RSPTA.212...75C}] 
which takes advantage of this new fiducial time for solar activity, we will not only see how solar, and solar-related, activity ``stacks up,'' we will identify a repeated pattern in the ocean index at those times indicating that there may indeed be a strong connection between the two systems on that timescale.
Correlation does not imply causation, but such a strong correspondence requires explanation, one that is beyond the current paradigm of atmospheric modeling.

\section{Results}

\begin{figure}[ht]
\centering
\includegraphics[width=0.75\linewidth]{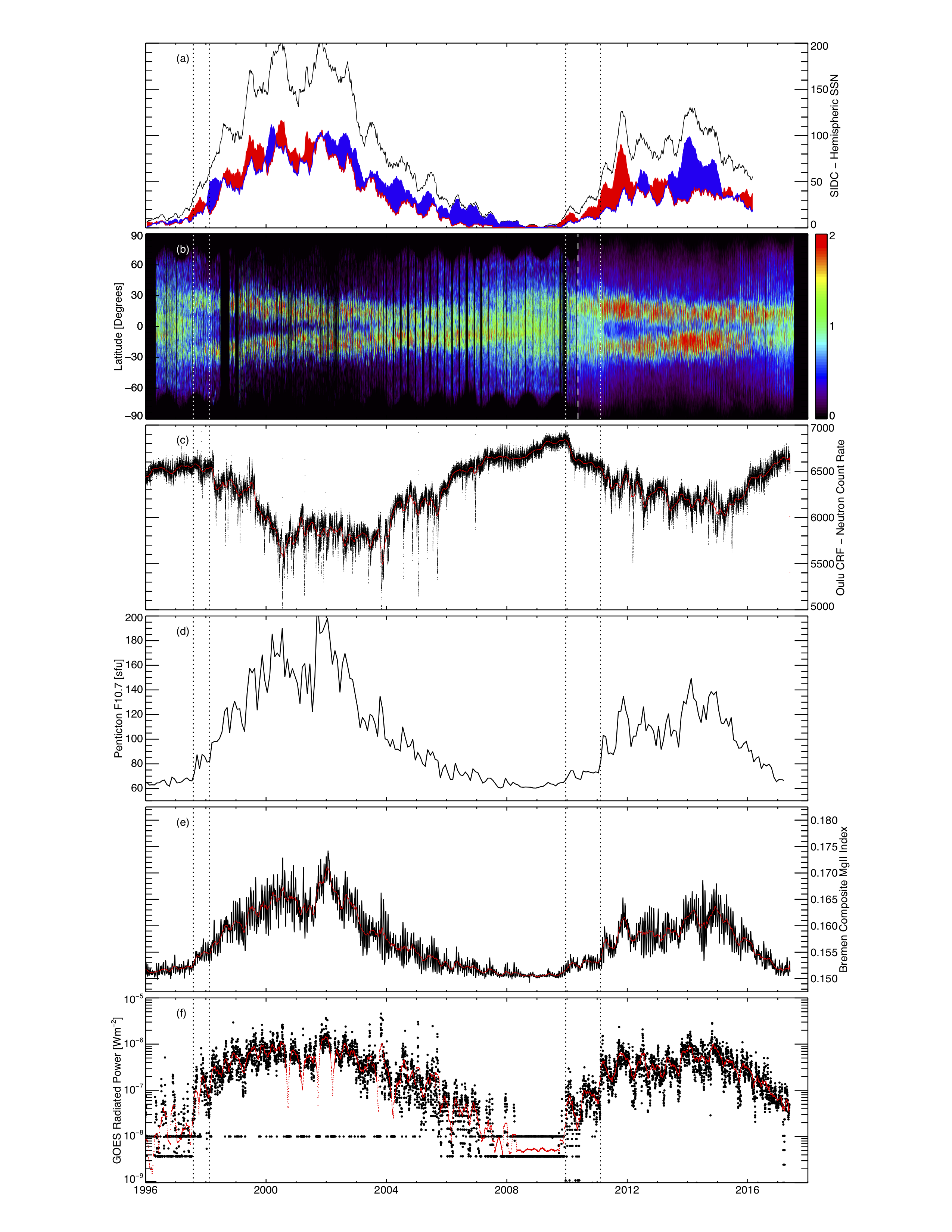}
\caption{Correlated variability of the Sun's output before and after the termination of solar cycles~22 and~23 in August of 1997 and February of 2011. From top to bottom: (A) the hemispheric sunspot number, as recorded by the Royal Observatory of Belgium; (B) the distribution of coronal EUV bright points as a function of latitude and time. (C) the cosmic ray flux as detected at Earth by the Oulu neutron monitor; (D) the Penticton 10.7cm radio flux; (E) the Mg~{\sc ii}\ index of ultraviolet variability from the University of Bremen; and (F) the integrated 1-8\AA\ X-Ray solar luminosity from the family of GOES spacecraft. In panel~(A) the red and blue traces correspond to the northern and southern numbers respectively; colored fill corresponds to a dominance of the corresponding hemisphere over the other. Throughout all panels, the apparent termination of the bands belonging to the 22-year solar magnetic activity cycle are marked with vertical dotted lines. These dates correspond to a sharp increase in sunspot activity in the northern hemisphere, spectral irradiance, and a decrease (and increased variability) in the cosmic ray flux. Notice that, during solar minima, the X-ray flux can fail to exceed the noise floor of the instrument. The dashed vertical white line in panel B represents the transition from SOHO/EIT to SDO/AIA data in May of 2010. }
\label{fig:f1}
\end{figure}

\subsection{Diagnostics Of The Cycle 22 and 23 Terminators}

Fig.~\ref{fig:f1} shows a combination of the primary measure used in M2019, the EUV Brightpoint (BP) density as a function of solar latitude with the variation of the Sun's hemispheric variability in spots (Panel A). The dramatic drop in BP density at the solar equator is visible in 1997 and 2011. Correspondingly, due to the higher quality data coming from the AIA instrument on the Solar Dynamics Observatory \cite{2012SoPh..275...17L} compared to its predecessor {SOHO/EIT} \cite{1995SoPh..162..291D}, the mid-latitude increases in activity beyond the 2011 termination. Progressing down the figure we see the anti-correlated variation of the galactic cosmic ray flux (CRF) as measured at the University of Oulu station (Panel C). The anti-correlation of CRF and solar activity \cite{1954JGR....59..525F,2013ApJ...765..146M}  is a result of changes in the Sun's global magnetic field strength (and structural configuration)---basically, a strong solar magnetic field blocks cosmic rays from entering the solar system, and hence the Earth's atmosphere with corresponding increases when said magnetic field is weak. Note that solar cycle 24 [ongoing from 2010] has seen a weaker global solar magnetic field than its predecessor [1996--2009]. From a radiative standpoint we show other canonical measures in the Penticton 10.7cm radio flux (Panel D); the composite index of the Sun's chromospheric variability measured through the ultraviolet emission of singly ionized Magnesium (Panel E),
a close proxy for solar ultraviolet flux at wavelengths near $\sim$200 nm that are important for molecular oxygen dissociation and ozone formation in the stratosphere; and the 1\--8\AA{} integrated coronal X-ray irradiance measured by the GOES family of spacecraft. Note that the final measure was the first in which terminator events were detected \cite{2005MmSAI..76.1034S,2009AdSpR..43..756S}.
In all of these cases the vertical dotted lines mark the termination points where step-function changes are present in each of the measurables (radiative increases and CRF decreases) that persist for the next several months.

\begin{figure}[ht]
\centering
\includegraphics[width=0.75\linewidth]{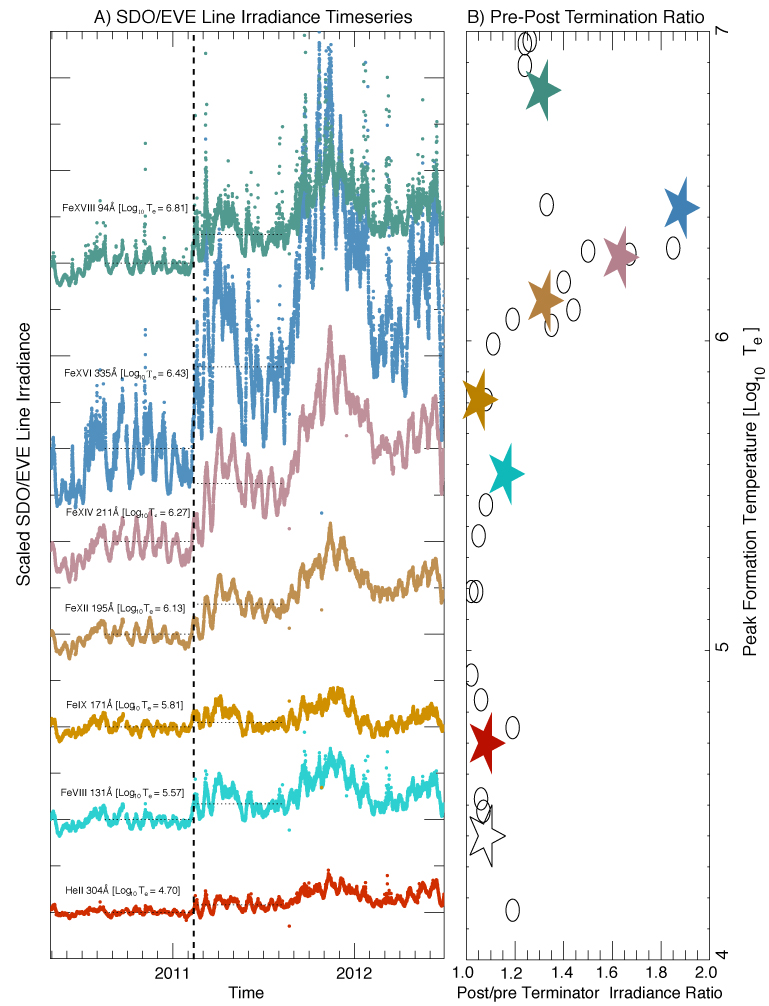}
\caption{The evolution of constituents of the  the Sun's spectral irradiance across the 2011 terminator as measured by SDO/EVE. (A) Time series of EVE emission of the seven AIA bandpasses, scaled to the time interval 180--60 days prior to the terminator. From bottom to top, the lines increase in their formation temperature. Each successively hotter line is offset on the $y$-axis by unity, except for the hottest line---Fe~{\sc xviii} 94\AA---which is offset by two to show the huge increase in emission in Fe~{\sc xvi} 335\AA. (B) The ratio of EVE emission across the terminator as a function of mean formation temperature for the spectral line. The step increase in emission is most pronounced from $\log_{10} T_e = 6.0$ to $6.4$; but even the hottest lines do show a greater increase than lines cooler than 1MK.}
\label{fig:f2}
\end{figure}

\begin{figure}[ht]
\centering
\includegraphics[width=\linewidth]{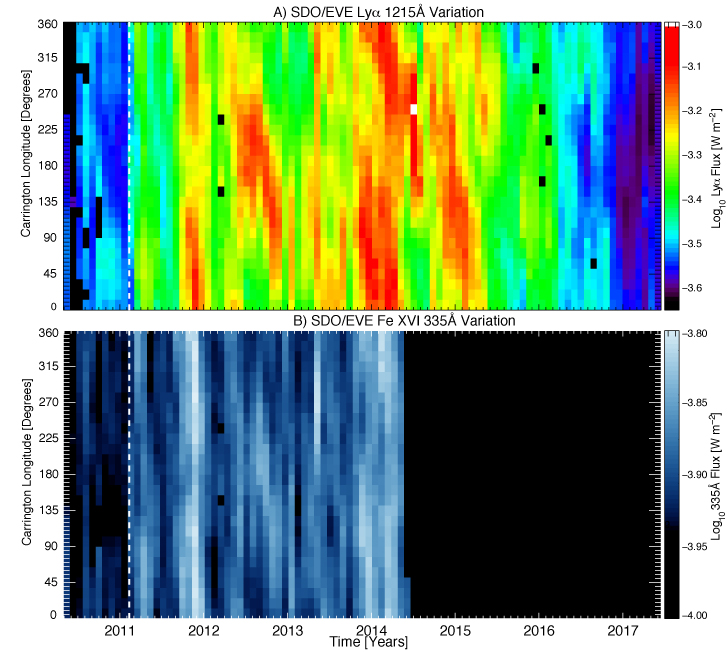}
\caption{Synthesizing the Longitudinal evolution of SDO/EVE measures of the Sun's spectral irradiance. (A)~Lyman-$\alpha$ emission from SDO/EVE as a function of Carrington Longitude and time for the whole SDO mission. The onset of emission at the terminator (dashed vertical line) in February 2011 is clear---and global.
(B)~Similarly, for Fe~{\sc xvi} 335\AA\ emission, which, per Fig.~\ref{fig:f2}(A), has the biggest step increase in emission over the terminator of the AIA lines. Note that the short-wavelength detector of EVE was only fully functional until May~2014.}
\label{fig:fS1}
\end{figure}

When assessing the impact of the Sun's variability on the Earth's atmosphere the primary culprit has been traditionally thought of as the solar cycle related changes in our star's spectral irradiance and its (clear) impact on the regions of the atmosphere about the stratosphere somehow coupling downward \cite{2010RvGeo..48.4001G}. The 2011 termination event allows us to observe the spectral variability of the event like never before---through the Sun-as-a-star measurements of SDO/EVE \cite{2012SoPh..275..115W}. 
Fig.~\ref{fig:f2} shows the variation in several EVE measures across the first two years of the SDO mission, including the 2011 termination. Arranged, from bottom to top, by  temperature of formation from (relatively) cool transition region emission in He~{\sc ii} (singly ionized Helium), to the hottest coronal emission of Fe~{\sc xviii} (seventeen times ionized Iron). The ratio of the pre- and post\--termination emission across that temperature range scales from 8\% to 85\% and is highly localized with plasma emission around 5 million Kelvin. This behavior has been noted also by two recent studies \cite{2017ApJ...844..163S,2017SciA....3E2056M}. Fig.~\ref{fig:f2} shows that highly optimized coronal emission starts immediately following Feb 11, 2011 (the black dashed vertical line). 
For contrast, in Fig.~\ref{fig:fS1}, we show the EVE data in a format to illustrate longitudinal behavior on the Sun in the 1215\AA{} (``Lyman $\alpha{}$'') and the Fe~{\sc xvi} 335\AA\ lines---at the peak of the emission increase shown in Fig.~\ref{fig:f2}. The timeseries of EVE data have been arranged in 27 day strips to approximate that of a complete solar rotation---a day of rotation relates to $\sim$13 degrees of longitude. In the cases shown (that bracket the range of plasma temperatures accessible to EVE) we see that, post\--termination, the activity of the Sun exhibits a global ``switch-on,'' that must be related to the global increase in magnetic flux emergence discussed by M2019.
We note that there is a second drop, of similar magnitude, in CRF in December 2009, that coincides with the increase (above the noise floor) in the GOES X-ray irradiance, and smaller increases in F10.7 and the Mg~{\sc ii} index.

To recap, the 2011 termination exhibits a $\sim$4\% decrease in the CRF and an (8--85\% from low to high temperature emission) increase in the ultraviolet photons that bathe our planet over {\em only} a few days.

\subsection{Terminators in Recent History}

\begin{figure}[ht]
\centering
\includegraphics[width=0.85\linewidth]{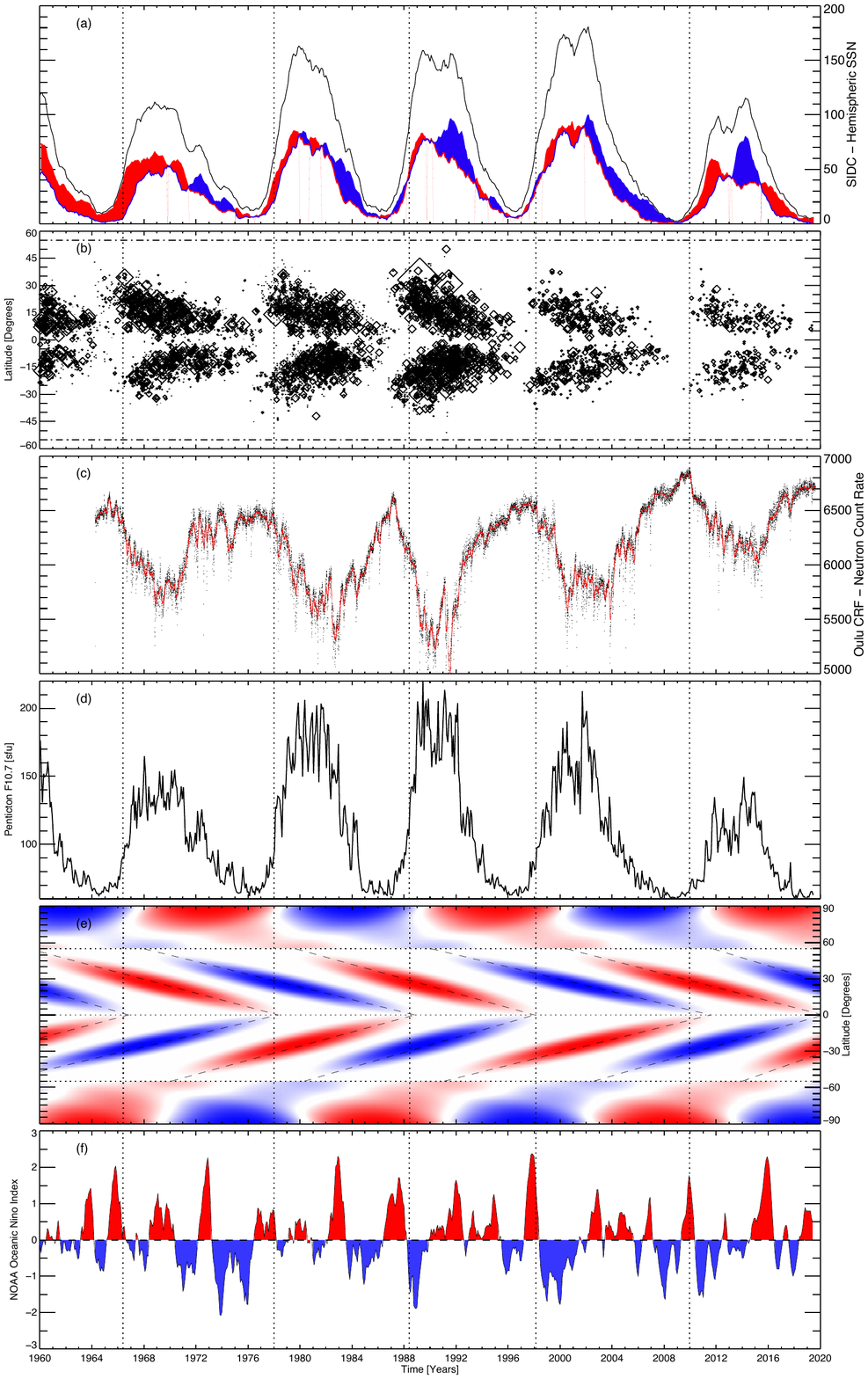}
\caption{Comparing more than five decades of solar evolution and activity proxies. 
From top to bottom: (A) the total (black) and hemispheric sunspot numbers (north---red, and blue---south); 
(B) the latitude--time variation of sunspot locations; 
(C) the Oulu cosmic ray flux; 
(D) the Penticton F10.7cm radio flux; 
(E) a data-motivated schematic depiction of the Sun's 22-year magnetic activity cycle; and 
(F) the variability of the Oceanic Nino Index (ONI) over the same epoch.
The black dashed 
lines mark the cycle terminators.
}
\label{fig:f3}
\end{figure}

Fig.~\ref{fig:f3} continues, and extends, our presentation of solar activity markers and proxies back over the past 60 years. We directly compare the variability of the total and hemispheric sunspot numbers with the latitudinal distribution of sunspots (the so-called ``butterfly'' diagram). Note that the terminator points, the family of vertical dashed black lines threading the panels of the plot (as developed in M2014) largely align with the very edges of the butterfly wings, noting that we do not use the symbol size in panel B to indicate the size of the spot---only that one was present. Panels C and D are extensions of those presented in Fig.~\ref{fig:f1} where the reader can appreciate the bracketing of the cycles provided by the termination points. Finally, panel E shows a data-motivated depiction of the latitudinal progression of the Sun's magnetic cycle bands. 
As initially developed by 
\citeA[hereafter M2014]{2014ApJ...792...12M}, 
these ``band-o-grams'' are set by three parameters (points in time): the times of hemispheric maxima (the time that the band starts moving equatorward from 55$^\circ$) and the terminator time. We assume a linear progression between those times in each hemisphere. 
Above 55$^\circ$ latitude we prescribe a linear 
progression of 10$^\circ$ per year, in keeping with ``Rush to the Poles'' seen in coronal green line data \cite{1997SoPh..170..411A}.
\citeA{2017NatAs...1E..86M} 
deduced that the temporal overlap and interaction between the oppositely polarized bands of the band-o-gram inside a hemisphere, and across the equator, was the critical factor in moderating sunspot production and establishes the butterfly diagram as a byproduct. The terminator is given as the time that the oppositely polarized equatorial bands cancel or annihilate and establish growth on the remaining mid-latitude bands. This gross modification of the Sun's global magnetic field has a impulsive growth on radiative proxies and a corresponding, inverse, relationship on the CRF as shown in Fig.~\ref{fig:f1}.

\subsection{Terminators and Oceanic Flips}
\label{sec:flip}

The bottom two panels of Fig.~\ref{fig:f3} compares the data-motivated band-o-gram with a measure of the El Ni\~{n}o Southern Oscillation (ENSO). There exist various indices to describe ENSO which include or exclude various components; we focus here on the NOAA-generated Oceanic Ni\~{n}o Index (ONI). Note that in this paper we are not trying to explore every bump and wiggle in the ONI---our primary focus are the ``decadal-scale''  large transitions from El Ni\~{n}o (hot mid-pacific) to La Ni\~{n}a (cold mid-pacific), the signature ``El Ni\~{n}o Events'' like that in 1997-98 \cite{2001JCli...14.1697T}. 
A visual comparison between the termination points in all panels and the ONI record of panel~(F) would appear to indicate that there is a possible relationship between them. 

To explore this potential relationship a little more we employ a modified version of the Superposed Epoch Analysis (mSEA) to the ONI, in addition to the solar activity measures presented above over the past sixty years with the termination points taken as the fiducial time. 
The methods behind the mSEA are described in Appendix~\ref{sec:msea}.

\begin{figure}[ht]
\centering
\includegraphics[width=\linewidth]{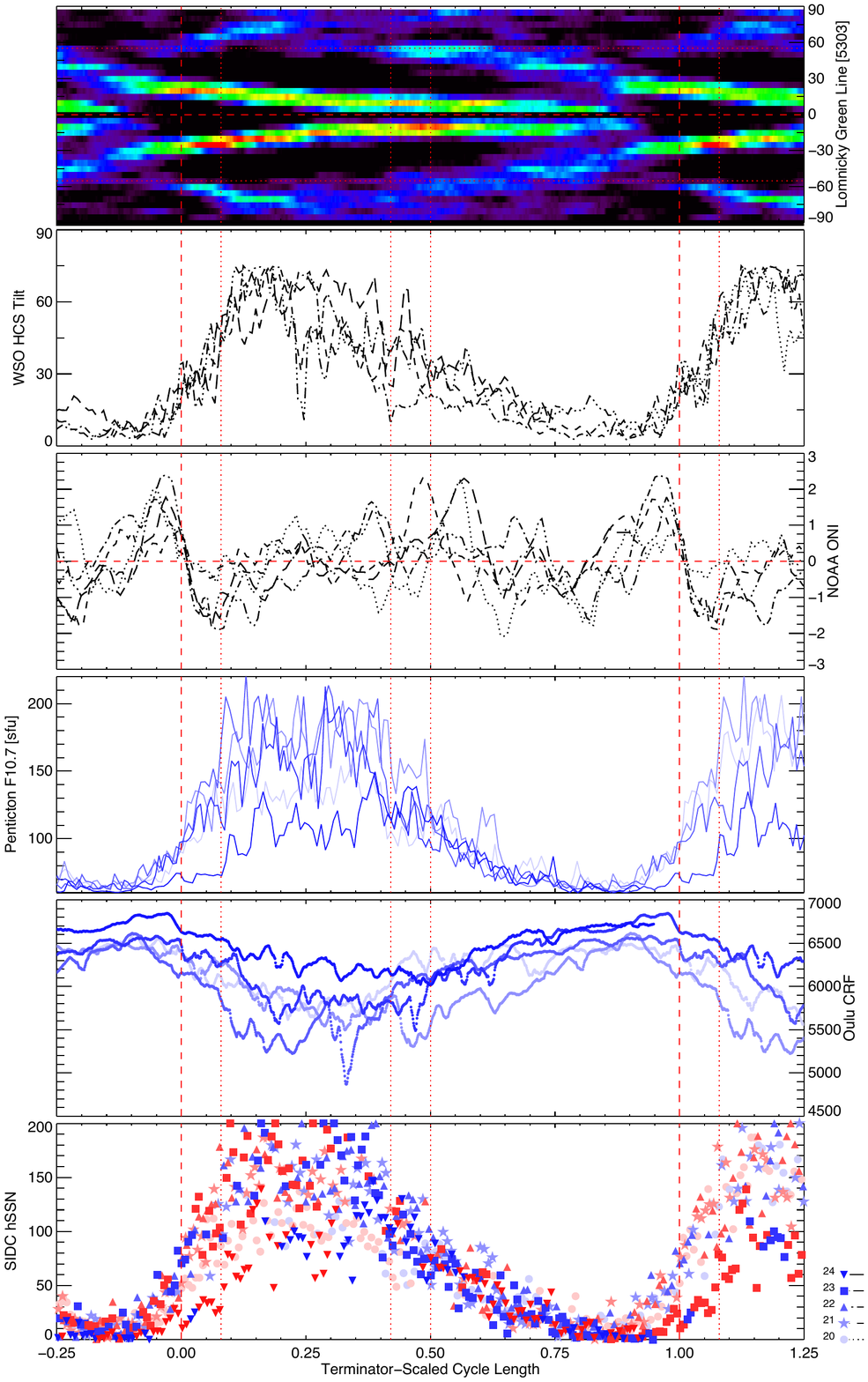}
\caption{
A modified superposed epoch analysis (mSEA) applied to the data shown in Figs.~\protect\ref{fig:f1} and~\protect\ref{fig:f3}.
{ 
The coherent behavior of all quantities (that is, more coherent than SSN) with the phase of the cycle is clear over cycles 20--24.}
The top two panels (new to this Fig.)\ show the computed tilt angle of the Heliospheric Current Sheet from Wilcox Solar Observatory and the NGDC Coronal Green Line data (replacing the schematic of Fig.~\protect\ref{fig:f3}E). 
The terminator signature is clearly simultaneously visible in solar, heliospheric {\em AND\/} atmospheric data. 
{The terminator is marked by the red dashed line at $x=0$ (repeated at $x=1$); 
a dotted line is placed purely to mark the half way point of the cycle at $x=0.5$, and 
two further dotted lines are placed at $x=0.08$ and $x=0.42$, marking the 
steep rise and fall-off
of the F10.7 flux (see text)---recall $x\sim 0.08$ is consistent with the period of a Rossby wave propagating in the solar tachocline \protect\cite{2017NatAs...1E..86M}. }
}
\label{fig:fS2}
\end{figure}

Fig.~\ref{fig:fS2} contains the 
key result of this paper.
It compares, from top to bottom, the mSEA of the maxima of Coronal Green Line emission \cite{1994SoPh..152..153R}, the computed Heliospheric Current sheet tilt \cite{1977SoPh...54..353S}, the ONI ENSO index, the Penticton 10.7cm radio flux, CRF, and the hemispheric sunspot numbers for the past six decades with respect to the termination points. (The last and first quarters of the cycle repeat to see the transitions across the terminator more clearly).   

To be open, the new data introduced in the top two panels do not overlap temporally with the other panels of Fig.~\ref{fig:fS2} (1939--1989 compared to 1965-present).
Nevertheless, as a composite ``standard cycle,'' 
using real data ({\em i.e.}, not a model or schematic),
they provides insight into the changes in, for example, F10.7 emission and GCR flux.
While there is considerable variability of the HCS in the declining phases of the solar cycle---owing, in part, to hemispheric asymmetry---the rise phase before and after the terminator is coherent. 
The bump at the terminator is real, and corresponds to the onset of the rush-to-the-poles emission above $55^\circ$ visible in the Green Line panel.

The clearest feature of the plot is that the mSEA indicates the ONI timeseries appears to collapse into a coherent regular behavior around the terminators:
a striking change from a strong El Ni\~{n}o to La Ni\~{n}a at the terminator. 
This would appear to indicate that in the depths of solar minimum conditions, when the radiative proxies are low and the CRF is high, there are epochs of warm Pacific conditions. 
Conversely, following the terminator, the rapid growth of radiative proxies and decline of the CRF would appear to systematically correspond to epochs of cooler Pacific conditions.

There is a general upward trend (with considerable scatter) in Pacific Ocean temperatures (increasingly positive ONI) as the solar activity cycle progresses.
Each of these phases last around $x=0.08$ in the normalized time scale, or about 10--11 months if the inter-terminator spacing is 11 years or so.
The $x=0.08$ scatter appears not inconsistent with the period of a Rossby wave  propagating in the solar tachocline \cite{2017NatAs...1E..86M}; clearly visible in the F10.7 flux panel of Fig.~\ref{fig:fS2}) or across the Pacific Ocean \cite{1979JPO.....9..663M}. 
However, near solar maximum a pattern returns:
three of the 5 cycles have a coherent second peak (El Ni\~{n}o) at around $x=0.5$, {\em i.e.}, a couple of years after sunspot maximum; the other two cycles, 22 and 23, have coherent double peaks around $x=0.33$ and $x=0.67$---recall that these are the two cycles with the largest offsets between maximum in the northern and southern hemispheres.

While there is small rise in F10.7 flux at the terminator, there is a consistently larger rise one tachocline Rossby period later. 
Indeed, 
we determine the peak rates of growth and decay of the (13-month smoothed) F10.7 flux for each cycle, and found that the edges where 
$|dF10.7 /dt| > 0.045  F10.7 $
are consistent with constant cycle phase $x=0.08$ and $x=0.42$; 
We thus plot the vertical dotted lines at $x=0.08$ and $x=0.42$.
After $x \sim 0.5$ there is very little F10.7 flux.
At the nadir of solar activity, as measured by either SSN or F10.7, we see the start of the coherent pre-terminator rise to El Ni\~{n}o in the ONI timeseries, while the GCR flux, which has been increasing since $x \sim 0.5$ (post-maximum), continues to rise until the next terminator.
Peak F10.7 emission (above $\sim$167 sfu), 
peak HCS tilt ($>60^\circ$) and 
peak sunspot activity 
are all well constrained by the dashed lines at 
$x = 0.08$ and 
$x = 0.42$,
corresponding to those phases of the cycle when there is significant Green Line emission above $55^\circ$, 
or when emission poleward of $55^\circ$ exceeds that of the new cycle branch that starts its equatorward journey from $55^\circ$.
Once the new cycle branch starts moving, the drop-off in all measures of solar activity, and increase in cosmic ray flux is clear and immutable.
Sunspot minimum occurs 
(at $x \simeq 0.8$)
when the four branches are of equal intensity \cite{2014ApJ...792...12M}.

Taken together, corpuscular radiation appears to have greater influence on ENSO than photons.

\section{Discussion}
\label{sec:disc}
%

In the previous section we have made use of a modified Superposed Epoch Analysis (mSEA) to investigate the relationships between solar activity measures and variability in a standard measure of the variability in the Earth's largest ocean---the Pacific. We have observed that this mSEA method brackets solar activity and correspondingly systematic transitions from warm to cool pacific conditions around abrupt changes in solar activity we have labeled termination points. These termination points mark the transition from one solar activity (sunspot) cycle to the next following the cancellation/ annihilation of the previous cycles' magnetic flux at the solar equator.

\begin{figure}[ht]
\centering
\includegraphics[width=0.80\linewidth]{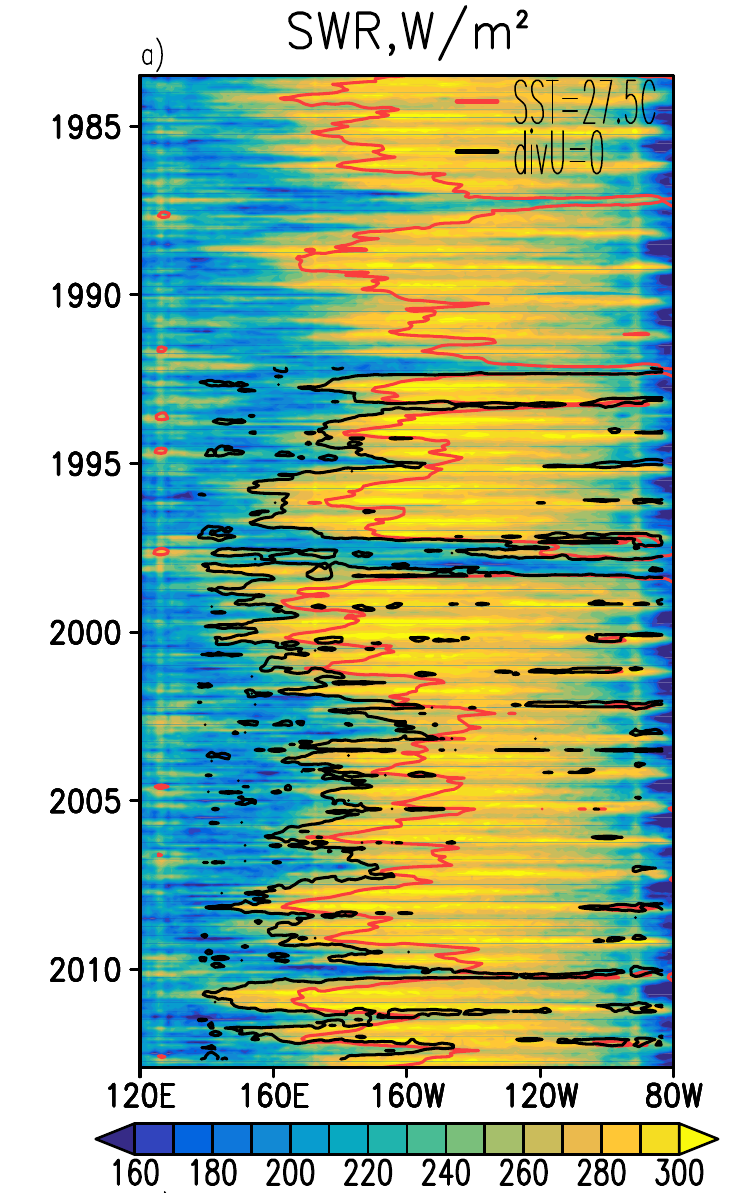}
\caption{
Time-longitude plot (Hovm\"oller representation) of anomalous incoming shortwave radiation (SWR in figure) along the equator in the Pacific sector. 
Overlain contours are the divergence of the surface zonal winds \mbox{(div $U=0$)} and the sea surface temperature 27.5$^\circ$C isotherm.
Reproduced ({\bf need permission!}) from \protect\citeA{2017JGRC..122.7880P}.
The difference in eastward excursions of the black and red contours between terminator-related ENSO events (extreme shifts beyond 120Win 1987-8, 1997-98 and 20107-98 and 2010)
and ``regular'' oceanic oscillatory patterns (rarely eastwards of 160W, if not the dateline)
is clear.
}
\label{fig:Pinker}
\end{figure}

Correlation does not imply causation however, the recurrent nature of the ONI signal in the terminator fiducial would appear to indicate a strong physical connection between the two systems. We do not present an exhaustive set of solar activity proxies, but it would appear that the CRF, as the measure displaying the highest variability, as something to be explored in greater detail in coupled climate system models. 
An overly simplistic explanation of the ONI evolution around terminators would appear to relate processes such as precipitation and cloud cover at times of high CRF with warming periods and the opposite at rapid reversals in those conditions. 
The correlation with short-term solar activity (Forbush decreases and sector boundary crossings) with weather---extra-tropical storm vorticity has been noted for several decades \cite{1973JAtS...30..135R,1989JGR....9414783T}.
It is not inconceivable, therefore, for the 3--5\%\ drop in GCR flux that occurs at a terminator, and that does not recover, 
may have a role in driving changes in large-scale weather patterns in the Pacific Ocean.

The rapidity in 
changes to cloud patterns is shown in 
Fig.~\ref{fig:Pinker}, which shows in a Hovm\"oller 
diagram (time-longitude) 
the
dramatic
eastward excursions of the black and red contours between terminator-related ENSO events (extreme shifts beyond 120$^\circ$W in 1987--98, 1997--98 and 2010).
We stress that this is {\em not\/} a manifestation of the 
Madden-Julian oscillation \cite{1972JAtS...29.1109M,2005RvGeo..43.2003Z}:
the difference between the dramatic eastward excursions and ``regular'' oceanic oscillatory patterns---rarely eastwards of 160$^\circ$W \cite{2005RvGeo..43.2003Z}, if not the dateline---is clear.
However, 
while there are many controversial aspects regarding possible effects of the MJO on ENSO \cite{2001BAMS...82..971Z},
the MJO may affect an El Ni\~{n}o by helping reduce the zonal gradient of sea surface temperature, or through exciting downwelling Kelvin waves of greater magnitude than typical reflection of Rossby waves at the Western boundary of the Pacific Ocean. 
\citeA{2005RvGeo..43.2003Z}, notes in his Section~6, and in particular the various McPhaden et al.\ references therein
\cite[for example]{1999Sci...283..950.}, 
``Extraordinarily strong MJO events have been repeatedly observed during the onset and growth stages of recent major ENSO warm events\ldots''; 
one may envision the nascent El Ni\~{n}o ``catching the MJO wave'' and riding it across the Pacific much as a surfer would closer to shore.

As discussed above, 
the solar cycle--QBO relationship is notoriously complicated, and atmospheric correlations with solar cycle variability are stronger if the data are sorted by east- or west-phase of the QBO. 
As such, it is beyond the scope of the present paper, 
but would be a fruitful exercise, 
to see how the timing of terminators with regards to the phase of the QBO affects the ENSO swing (recall the downward propagation of westerlies is typically much smoother than that of easterlies), or indeed if the timing of terminators affects the phase of the QBO.
Or given there are 4--5 QBO flips per solar cycle, we  
need to look at solar variability at similar timescales, {\em i.e.}, the solar Rossby wave driven activity surges 
which are of comparable magnitude to the entire solar cycle variation in spectral irradiance
\cite{2017NatAs...1E..86M}.
The strong El Ni\~{n}o of 2015--2016 undoubtedly played a role 
(together with a warmer than average troposphere and cooler than average stratosphere)
in the anomalous rising westerlies in the QBO those years \cite{2016GeoRL..43.8791N}.
As of late 2019, there is not yet evidence of bifurcation in the QBO with continued tropospheric warming, stratospheric cooling, and reduced Walker circulation, even though the predicted El Ni\~{n}o 
(see Section~\ref{sec:slab2}, and Appendix~\ref{sec:unit})
endures.

\subsection{Potential Mechanisms}
\label{sec:GCR}

\citeA{1978JATP...40..121H} argued that
Galactic cosmic ray decreases tend to enhance the electric field at low heights. The protons produce excess ionization near and above 10--20 km, resulting in maximum changes in temperature and ionization rates for a Forbush decrease at altitudes typical of cirrus formation, and greatly increasing the atmospheric conductivity and possibly lowering the height of the ``electrosphere.'' 
This is also the height region where electrical conductivity is lowest, so if electric fields are involved in the cloud microphysics \cite{2010RvGeo..48.4001G,2000SSRv...94..231T,Harrison2004} only small amounts of energy are needed to initiate changes in cloud properties. 
Consequent effects near the solar proton cut-off latitude also lead to an enhancement of the atmospheric electric field near the surface. If appropriate meteorological conditions (warm moist air with updrafts) exist or develop during a solar event, the atmospheric electric field enhancement may be sufficient to trigger thunderstorm development. 

Extending throughout the atmosphere from the Earth's surface to the lower ionosphere, the global atmospheric electric circuit provides a conduit, via the solar modulation of cosmic rays and resulting ionization changes, for a solar influence of meteorological phenomena of the lower atmosphere. 
Other possibilities for the coupling of GCRs to storm vorticity, at least for Forbush decreases, and at higher latitudes, include 
atmospheric gravity waves propagating from the auroral ionosphere \cite{2009AnGeo..27...31P}
and
precipitation of relativistic electrons from the radiation belts \cite{2012AdSpR..50..783M}.
%
An analysis of Radiation Environment Monitor (IREM) housekeeping data from 2002--2016 on the INTEGRAL Gamma Ray Observatory, albeit one focused on the vulnerability of operating spacecraft to so-called ``killer'' electrons \cite{2017SpWea..15..917M},
strongly suggests the latter 
through a step change in 1.27 MeV electron flux coincident with the step change in GCRs  
seen in Figs.~\ref{fig:f1}, \ref{fig:f3} and~\ref{fig:fS2}.
Similar results from NOAA~POES data were 
recently
reported by \citeA{doi:10.1029/2018JA025890}, who then correlated the precipitation of energetic electrons with an Antarctic coherent spaced-antenna wind profiler and found that \mbox{$>30$~keV} electrons penetrated the stratopause ($\sim$55~km).

An alternative explanation is that changes in the CRF alone drive the ENSO flip by ionization seed particle formation and subsequent effects on aerosol processes \cite{1997JASTP..59.1225S,2002GeoRL..29.2107K,2016JGRA..121.8152S,2017NatCo..8E2199M}.
However, the effects of cosmic rays on cloud formation are a matter of hot debate \cite{ROG:ROG1696,2017JGRD..122.8051P}, with even the sign of the correlation between cosmic rays and climate not agreed on---for enhanced low-altitude cloud the dominant effect would be reflection of incoming shortwave solar radiation (a cooling effect); conversely for enhanced high-altitude cloud, the dominant effect would be the trapping of re-radiated, outgoing longwave radiation (a warming effect). 
Some studies have found no statistically significant correlations between the CR flux and global albedo or globally averaged cloud height \cite{krissansentotton2013}, but most such studies focus on the immediate cloud cover effects of Forbush decreases 
({\em i.e.}, the Cosmic Ray Flux decrease associated with a CME).
which drop suddenly, but then recover on a timescale of a day or days. 
It is not surprising then that there is little evidence that these events are apparent in cloud data sets, compared to the sustained drop of 3--3.5\% in GCRs following terminators that does not recover. 
Rather than direct cloud cover, 
\citeA{1973JAtS...30..135R}  
showed a correlation between geomagnetic storms and increased storm vorticity over the Northern Pacific ocean.
\citeA{1989JGR....9414783T} extended the earlier storm vorticity analysis, again looking at superposed epochs, but for Forbush decreases
(rather than than the associated geomagnetic storms).
\citeA{2014AdSpR..54.2491A}
similarly studied Forbush decreases and large-scale (some $15^\circ$--$30^\circ$ in longitude) mid-latitude atmospheric pressure variations, with the strongest correlations over Europe and European Russia, and conjugate southern hemisphere areas.
Both \citeA{1973JAtS...30..135R} 
and
\citeA{2014AdSpR..54.2491A}
also discussed the possible amplifying effects of cloud microphysical processes, including the electric field motivation discussed above \cite{2000SSRv...94..231T,Harrison2004}.

Independent of the exact mechanisms of coupling solar modulation of GCRs to ENSO, which are beyond the scope of this manuscript,
the
results discussed above and shown in Fig.~\ref{fig:fS2} hold for the past 5 solar cycles, or 60 or so years. 
Cosmic ray data do not, unfortunately, extend back to the preceding terminator.
The question must be asked, then, why has the regular pattern of Fig.~\ref{fig:fS2} occurred and reoccurred regularly since 1966?

\subsection{Atmospheric Changes}
\label{sec:AGW}

{\color{blue}

Recent 
simulation runs
suggest
changes in 30-year cloud and net feedback are a plausible answer \cite{2016NatGe...9..871Z}, given that the extremum feedback (closest to zero; all values are negative) occurred in 1945, and has been increasingly negative ever since.
Similarly, 
a $\sim$4--6\%\ decrease in cloud cover over the western Pacific ($\sim$140--160$^\circ$E) has been reported from the 
Extended Edited Cloud Reports Archive (EECRA) 
ship-borne observations since 1954 \cite{2014JCli...27..925B}, and a comparable increase over the mid Pacific ($\sim$150--120$^\circ$W).
160$^\circ$E is the ``balance point'' about which
SSTs flip-flop in an El Ni\~{n}o--La Ni\~{n}a transition 
\cite{2017JGRC..122.7880P}. 
Thus over the past several decades the cloud pattern in the western Pacific has adopted an almost El Ni\~{n}o-like default state, 
consistent with an observed eastward shift in precipitation in the tropical Pacific and weakening of the Walker circulation over the last century \cite{2004JCli...17.3109D,2007JCli...20.4316V},
and which has been tied, via simple thermodynamics, to a warmer atmosphere.
Even allowing for model uncertainties, the 
simplest explanation is that the
weakened Pacific Walker circulation and less cloudy Western Pacific enables the relatively constant terminator-driven changes in CRF (and thus clouds) to have sufficient ``impact'' to flip the system from El Ni\~{n}o to La Ni\~{n}a, 
independent of the actual mechanism that couples CRF changes to clouds and ENSO.
Not unrelated, other
consequences of a reduced Pacific Walker circulation include an increase in tropical Atlantic vertical wind shear \cite{2007GeoRL..34.8702V}, 
which, topically, 
modulates the intensity of the Atlantic hurricane season.

} 

It is probably not a coincidence that the increasingly negative feedback between cloud cover changes and net irradiance since WW2, and the decreased western Pacific cloud cover
corresponds to the close-to-monotonic rise in sea surface temperatures over the same time period (the ``hockey stick'' graph).
Tropospheric warming leads to stratospheric cooling
\cite{Ramaswamy1138};
does the changed stratosphere make it more susceptible to amplifying transient changes in CRF to phase changes in QBO?

Evidence for a changing Brewer-Dobson circulation---the mass exchange between troposphere and stratosphere characterized by persistent upwelling of air in the tropics---comes from satellite and radiosonde data, which indicate a reduction in temperatures and ozone and water vapor concentrations over the past four decades, particularly in the tropical lower stratosphere at all longitudes 
\cite{2005JCli...18.4785T},
pointing to an accelerated tropical upwelling 
\cite{2008JGRD..113.6107R}.
\citeA{2019RvGeo..57....5D} 
discuss the teleconnection of ENSO both vertically, to the stratosphere, and thence latitudinally affecting the strength and variability of the stratospheric polar vortex in the high latitudes of both hemispheres:
El Ni\~{n}o events are associated warming and weakening of the polar vortex in the polar stratosphere of both hemispheres, while a cooling can be observed in the tropical lower stratosphere. 
These impacts are linked by a strengthened Brewer-Dobson circulation,
with planetary waves generated by latent heat release from tropical thunderstorms being the likely modulation mechanism 
\cite{DeckertDameris08,2019RvGeo..57....5D}. 
Such circulation changes are
only likely to intensify in a future with higher tropical heat and moisture at the sea surface, affecting not only tropospheric climate but also stratospheric dynamics,
increasing the likelihood of the future terminator-driven changes in CRF having sufficient ``impact'' to flip the system from El Ni\~{n}o to La Ni\~{n}a.

{\color{blue}

It is beyond the scope of the present paper, 
but would be a fruitful exercise, 
to see how the timing of terminators with regards to the phase of the QBO affects the ENSO swing (recall the downward propagation of westerlies is typically much smoother than that of easterlies), or indeed if the timing of terminators affects the phase of the QBO.
The strong El Ni\~{n}o of 2015--2016 undoubtedly played a role 
(together with a warmer than average troposphere and cooler than average stratosphere)
in the anomalous rising westerlies in the QBO those years \cite{2016GeoRL..43.8791N}.
Since we may predict enduring El Ni\~{n}o-like conditions through 2019 
(see Section~\ref{sec:slab2}, Appendix~\ref{sec:unit}), 
might we again see bifurcation in the QBO with continued tropospheric warming, stratospheric cooling, and reduced Walker circulation?  
Similarly 
it is beyond the scope of this paper, 
but also plausible, 
that the annual or inter-annual variations in ENSO  (see, {\em e.g.}, within the extended double-dip La Ni\~{n}as of 1999--2002, or 2010--11 \cite{2012AusBOM..LaNina}) are driven by annual or inter-annual variations in solar activity \cite{2015NatCo...6E6491M,2017NatAs...1E..86M}.
Similarly, the difference between ``canonical'' and ``Modoki'' (or Eastern and Central Pacific) flavors of El Ni\~{n}o \cite{2001JCli...14.1697T,2009Natur.461..511Y,2009JCli...22..615K}
may be
associated with
the difference between those forced by ``fixed'' solar cycle landmarks and those responding to oceanic/ atmospheric dynamics, 
 although increasingly warming sea surface temperatures may just be the dominant Modoki El Ni\~{n}o driver \cite{2009Natur.461..511Y}.
 
} 

\subsection{Future Predictions}
\label{sec:slab2}

As discussed in \citeA{2014ApJ...792...12M}, the band-o-gram developed
therein could be extrapolated linearly out in time---indeed the versions
shown here (e.g., Fig.~\ref{fig:f3}) extend to 2020. 
The linear extrapolation of the solar
activity bands outward in time was verified in 
\citeA{2017NatAs...1E..86M}
by updating the original analysis and comparing to the earlier
band-o-gram. M2014 projected that cycle~25 sunspots would start to
appear in 2019 and swell in number following the terminator in early
2020. 
That projection appears to still hold. 
Based on the mSEA of the past sixty years we may therefore expect
that in 2019 we will see a 
{
persistent (two-year)} 
El Ni\~{n}o event which is followed by
a rapid transition into La Ni\~{n}a conditions in 2020 following the
terminator between sunspot cycles~24 and~25.

We do not attempt to predict the intensity of either the coming El
Ni\~{n}o or La Ni\~{n}a, merely the timing with regards the impending
terminator. Indeed, this prediction does require the caveats that:
({\em i}) 
there is more to ENSO effects than the single variable ONI;
and
({\em ii}) 
the outcomes of each ENSO event are never exactly the same: they depend
on the intensity of the event, the time of year when it develops and the
interaction with other climate patterns extant in the atmosphere (the
phase of the QBO, for example).
Nevertheless, the correlation with ONI and terminators over sixty years 
strongly favors a repeat in 2019-20.

No obvious relationship between the summer or winter dates in
Table~\ref{tab:t1} and the strength of either the El Ni\~{n}o preceding the
terminator or the La Ni\~{n}a following.
But it might be plausible that 
an event
happening in boreal summer will affect the atmospheric and
oceanic responses differently than if it happened in austral summer due
to the interannual variability of CO$_2$ concentrations 
(which, as measured at Mauna Loa, peak in May, and are driven primarily by seasonal changes in northern hemisphere foliage cover).
Whether the impact of such variation is greater or lesser than  
the phase of the QBO at the Terminator remains for future investigations.

Appendix~\ref{sec:unit} extends the prediction for the whole of solar
cycle~25, based on the band model predictions of the terminators for
cycle~24 in April 2020, and that of cycle~25 in October~2031 (with an
error bound of $\pm 9$ months). El Ni\~{n}os may be expected around 2026
and 2031, and La Ni\~{n}as in 2020--21, 2027--28 and 2032--33. Near the
(sunspot) maximum of cycle~25, we expect predominantly neutral
conditions.
Further, given the relative strength of Atlantic hurricane seasons in the first year of La Ni\~{n}a after an El Ni\~{n}o, when waters are still warm but upper level wind shears are favorable for cyclone genesis, we may expect a particularly active season in 2020 or 2021.

\subsection{Wavelet Analysis}
\label{sec:wavelet}

\begin{figure}[ht]
\centering
\includegraphics[width=\linewidth]{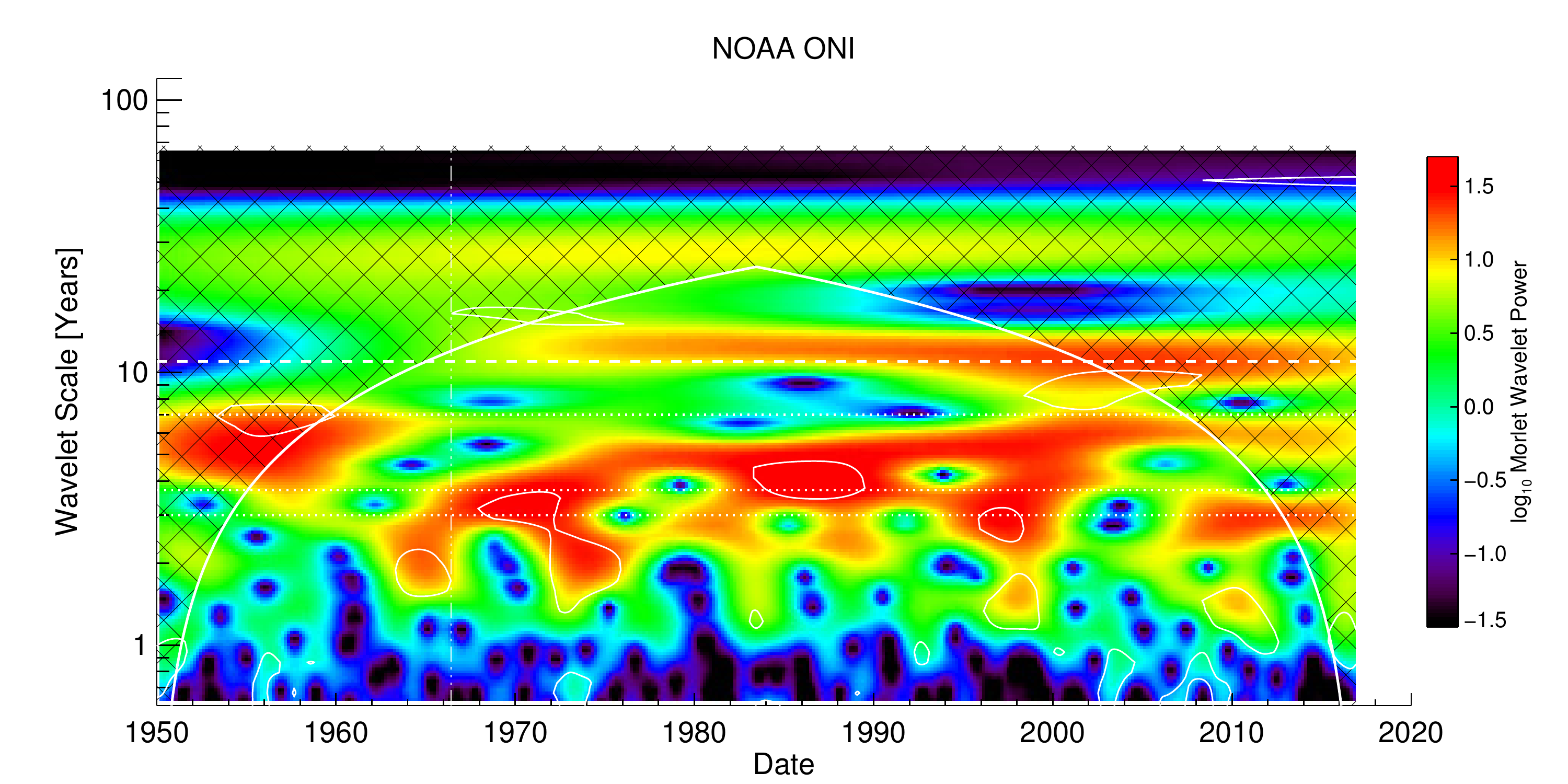}
\includegraphics[width=\linewidth]{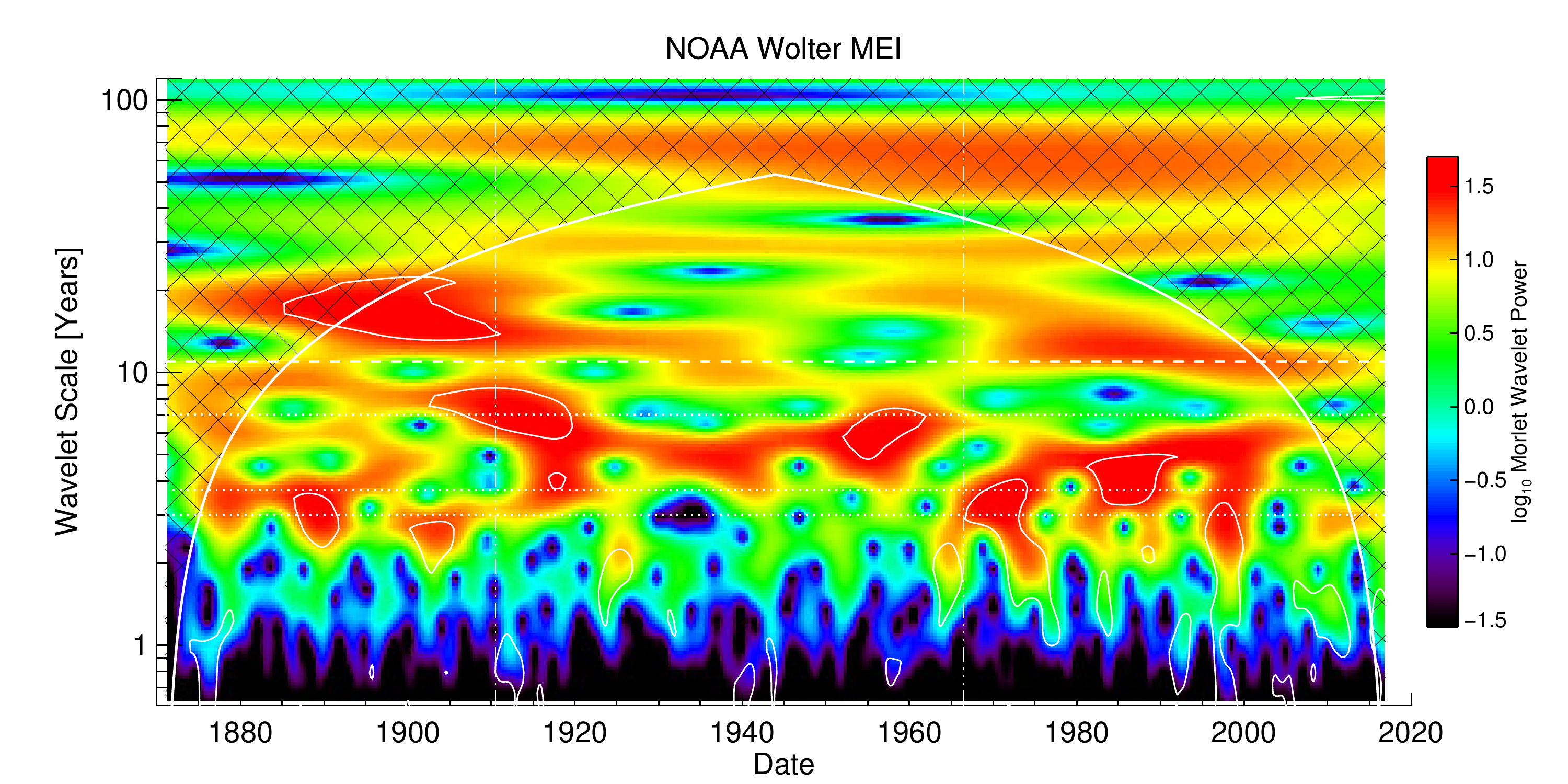}
\caption{Wavelet power spectra for the NOAA indices ONI 1950--present (top) and the extended ``Multivariate ENSO Index'' (MEI) 1871--present (bottom; note change of abscissa scale). 
In each panel, the contours enclose regions of greater than 90\% confidence for a red-noise process.
Cross-hatched regions on either end indicate the “cone of influence,” where edge effects become important. 
Horizontal dashed and dotted white lines refer to periods of 3, 3.7, 7, and 11 years;
Vertical white lines indicate June~1966 (the Cycle~19 terminator), 
and, in the MEI panel, January~1911 (see text).
Significant power is seen at solar cycle scales from the mid-1960s on, 
consistent with the results of \protect\citeA{TorrenceCompo98}, and \protect\citeA{1996JCli....9.1586W}.
}
\label{fig:wavelet}
\end{figure}

{
Given that the key result of the present paper is that ENSO variability is correlated with the terminators, 
which occur not a fixed temporal frequency but at a fixed phase of the solar cycle,
we are reticent to include a Fourier spectral analysis.
Nevertheless, the question ``Would you expect there to be significant power in a Fourier spectrum of the entire ENSO signal?''
is a valid one, as there have been several previous spectral analyses of ENSO.
Indeed, the
seminal wavelet analysis paper \cite{TorrenceCompo98} uses ENSO data (the Ni\~{n}o3 timeseries) as its ``practical example.''

As such, Fig.~\ref{fig:wavelet} shows wavelet power spectra for the ONI index as discussed above, and also for the longer term ``Multivariate ENSO Index,'' MEI, 
\cite{2011IJCli..31.1074W}
that combines air pressure, temperature and wind speed data along with sea surface temperatures, normalized such that the mean value for 1871--2005 is zero and the standard deviation is unity.  

As a sanity check, the spectra of the two indices agree, and our analysis agrees with previous ENSO wavelet analyses \cite{1996JCli....9.1586W,TorrenceCompo98} that
``the principal period of ENSO has experienced two rapid changes since 1872, one in the early 1910s and the other in the mid-1960s.''
Thus in both panels of Fig.~\ref{fig:wavelet}, vertical dot-dashed lines indicate June~1966 (the cycle~19 terminator), and in Fig.~\ref{fig:wavelet}b, somewhat arbitrarily, January~1911 marking the extent of the significance contour at 12--14 year scales and low power at scales shorter than about 4 years. 
An abrupt alteration
anywhere between 1911 and 1914 would not be inconsistent with Fig.~\ref{fig:wavelet}b.
However, given 
the likely role of tropospheric warming and stratospheric cooling in changing the properties of ENSO \cite{Ramaswamy1138}, and polar vortex--QBO teleconnections 
\cite{1988JATP...50..197L,2014ACP....1413063T},
it is believable 
that the June 1912 Novarupta volcano eruption in Katmai National Park, Alaska \cite{Fierstein1992}---the largest eruption of the 20th century in terms of ash volume expelled, and which, 
unlike other major eruptions with stratospheric con\-sequences, happened at high rather than equatorial latitudes---could be the trigger of the 1910s phase change seen in Fig.~\ref{fig:wavelet}.
Another suggestion from Fig.~\ref{fig:wavelet} 
is
that another abrupt alteration of oscillation period occurred around 2003--5 to a dominant 3-year periodicity. 
Even though one could then argue that a 3-year intrinsic periodicity would also make a 2019--2020 prediction,
the power at scales of a few years (almost always) exceeds that at solar cycle scales, 
and 
there is consistent, significant, power at 11-ish year scales over the past five solar cycles. 

Not unrelated to the change in ENSO principal period and the onset of a significant signal at solar cycle scales
in the mid-1960s,
\citeA{Wang95} noted that the onset of El Ni\~{n}o experienced an abrupt change in the late 1970s. He attributed the change to ``a sudden variation in the background state, associated with `a conspicuous global warming' and deepening of the Aleutian Low in the North Pacific.''

Finally,
we reserve a fuller treatment of the North Atlantic Oscillation (NAO; essentially the atmospheric pressure difference between Iceland and the Azores) for a future paper, but we will mention here that a similar terminator-NAO correlation exists, specifically a transition from a negative to a positive phase of NAO. 
This is reasonably well understood in that the (post-terminator) surge in EUV photons gives rise to anomalous heating of the equatorial upper stratosphere that alters the (winter) stratospheric circulation, coupling downwards to the troposphere at higher latitudes, flipping the phase of the NAO
\cite{2002JGRD..107.4749K,2010RvGeo..48.4001G}.

} 

\section{Conclusion}
\label{sec:dropmike}

We have shown a strong correlation between solar and tropospheric variability, in that swings from El Ni\~{n}o to La Ni\~{n}a
are related to the phase of the solar cycle's ``fiducial clock,'' 
and 
that that clock does not run from the canonical solar minimum or maximum, but instead resets when all old cycle flux is gone from the solar disk.
{While the exact mechanism remains to be elucidated, changes in cosmic ray flux appear to the be the driver of these ENSO swings.}

Finally, in the absence of sensitivity to solar-driven cosmic ray flux variations in current coupled climate models, we have less than a year or so to wait to see if this indicator pans out. 
However, should the coming terminator be followed by such an ENSO swing then we must seriously consider the capability 
of coupled global terrestrial modeling efforts to capture ``step-function'' events, and assess how complex the Sun-Earth connection is, with particular attention to the relationship between incoming cosmic rays and clouds/ precipitation over our oceans. 
The challenge then has to be
``what needs to be done to
disprove the causal link between terminators and ENSO?''

\acknowledgments
R.J.L.\ was supported by an award to NASA GSFC from the NASA Living With a Star program. 
The National Center for Atmospheric Research is sponsored by the National Science Foundation and the compilation of feature databases used was supported by NASA grant NNX08AU30G. 
All ground-based and spacecraft data presented here are publicly available from their respective observatories or archivers. 
We thank Lesley Gray, Justin Kasper, Kevin Trenberth, Nicholeen Viall and Sandra Chapman for constructive discussions 
improving 
earlier versions of
the manuscript.


\appendix
\section{The Coronal Green Line and Extended Solar Cycle}
\label{sec:ngdc}

Starting with the advent of the coronagraph in the late 1930s, routine measurements were made of the 5303\AA\ ``green line'' of the corona, even before it was known that it was emission from highly ionized iron (Fe~{\sc xiv}).
Multiple researchers, including 
\citeA{1988Natur.333..748W}, 
and
\citeA{1997SoPh..170..411A,2003SoPh..216..343A}, 
showed that the intensity at high latitudes ($>60^\circ$, or at the very least, higher than the highest observed sunspots) manifested an ``extended'' solar cycle.
Further, the high-latitude emission was situated above the high-latitude neutral line of the large-scale photospheric magnetic field, thus implying a connection with the solar dynamo.

All in all, the 5303\AA\ Green Line observations are an extended duration record providing evidence for an ``extended'' solar cycle that begins every 11 years but lasts for approximately 19–20 years.
The band-o-gram of Fig.~\ref{fig:f3} implicitly assumes this Wilson-like 19-20 year progression from $55^\circ$ to equatorial termination.

Fig.~\ref{fig:fS2} replaces the band-o-gram of Fig.~\ref{fig:f3} with the computed HCS tilt angle from the Wilcox Solar Observatory \cite{1977SoPh...54..353S}  for the 3.5 cycles that data has been extant, and the NGDC composite Coronal Green Line data \cite{1994SoPh..152..153R}.
To be clear, the data in this panel does not overlap temporally completely with the other panels of Fig.~\ref{fig:fS2} (1939--1989 compared to 1965-present).
Nevertheless, as a composite ``standard cycle,'' it provides insight into the changes in, for example, F10.7 emission and GCR flux.
For clarity, we track the local maxima in emission, allowing 4 per hemisphere, and plot the ``average track'' of peak emissions, as a function of time, and then compile them in an mSEA analysis (see Appendix~\ref{sec:msea}).
And
for completeness, we have remade the Green Line composite mSEA using SSN max and min as the fiducial time; the spread of peak intensity tracks 
is optimum using the terminators, as shown in Fig.~\ref{fig:fS2}.

\section{Modified Superposed Epoch Analysis}
\label{sec:msea}

The concept of a Superposed Epoch Analysis [SEA] 
was originally conceived 
(appropriately enough)
by \citeA{1913RSPTA.212...75C}
for the purpose of correlating sunspots with terrestrial magnetism---the recurrence in geomagnetic data of the 27-day Carrington periodicity.
Similar techniques were used by \citeA{1973JAtS...30..135R}  
to show the correlation between geomagnetic storms and increased storm vorticity over the Northern Pacific ocean.
\citeA{1989JGR....9414783T} extended the earlier storm vorticity analysis, again looking at superposed epochs, but for Forbush decreases
({\em i.e.}, the Cosmic Ray Flux decrease associated with a CME, rather than than the associated geomagnetic storms.
\citeA{1973JAtS...30..135R} also discussed the possible amplifying effects of cloud microphysical processes.
Recall from Figs.~\ref{fig:f1} and~\ref{fig:f3} that a 3--5\%\ drop in GCR flux that occurs at a terminator, which that does not recover; it is not inconceivable that such a change is responsible for changes in large-scale weather patterns in the Pacific Ocean.

Rather than defining a standard superposed epoch analysis repeating over some number of days/years, the critical modification here is to first scale time to be fractions of a cycle, from terminator to terminator.
(One may think of this then as a ``phase'' of the solar cycle, but we choose here to express length in terms of a fraction 0--1 rather than 0--$2\pi$.)

\begin{table}[t]
\centering
\begin{tabular}{|l|l|r|}
\hline
Cycle & Terminator Date &  mSEA Temporal \\
{} & (Observed) & Shift (Days) \\
{} & (Predicted) & {} \\
\hline
19 & 1966 June 01 & 0 \\
\hline
20 & 1978 Jan 01 & 0 \\
\hline
21 & 1988 June 01 & $-100$  \\
\hline
22 & 1997 Aug 01 & $+30$ \\
\hline
23 & 2009 Dec 15 & $+100$ \\
\hline
24 & 2020 Apr 01\ldots? & \ldots \\
\hline
\end{tabular}
\caption{\label{tab:t1}
Temporal shifts applied to EUV BP Terminator dates to align to step changes in GCR record. See Figs.~\protect\ref{fig:f3} and~\protect\ref{fig:fS2}. 
}
\end{table}

The terminator dates are defined from the band-o-gram (solar data), 
but,
motivated by \citeA{1973JAtS...30..135R} and \citeA{1989JGR....9414783T},
are then adjusted manually by up to $\pm100$ days such that the GCR traces line up.
Table~\ref{tab:t1} shows a list of temporal shifts applied.
So as Fig.~\ref{fig:fS2} shows, even though everything is tied to GCR, there is tight correlation between cycle timings in F10.7 especially and ENSO. 
(Recall that the cadence of the ENSO data in Figs.~\protect\ref{fig:f3} and~\protect\ref{fig:fS2} is only monthly.)

Expressing cycle progression as a fraction of their length requires the terminator of cycle 24 to be hard-wired. It is set to be April 1, 2020, based on our current extrapolation of the equatorial progression of EUV BPs \cite{10.3389/fspas.2017.00004}.
If Cycle 24 doesn't terminate until later, then the corresponding traces in Fig.~\ref{fig:fS2} would be compressed leftwards; however, an inspection of the rising and falling edges of the F10.7 trace does suggest that mid-2020 is an accurate assumption at the time of writing.
This date implies that Cycle 24 is relatively short, at less than 10.5 years, compared to over 12 years for Cycle 23.

\section{Statistical Tests}
\label{sec:stat}

\subsection{Terminators and ENSO: Monte Carlo Simulations}
\label{sec:monte}

We can quantify the apparent correlation between terminators and ENSO crossings by employing three different Monte Carlo simulations.

\begin{figure}[ht]
\centering
\includegraphics[width=0.475\linewidth]{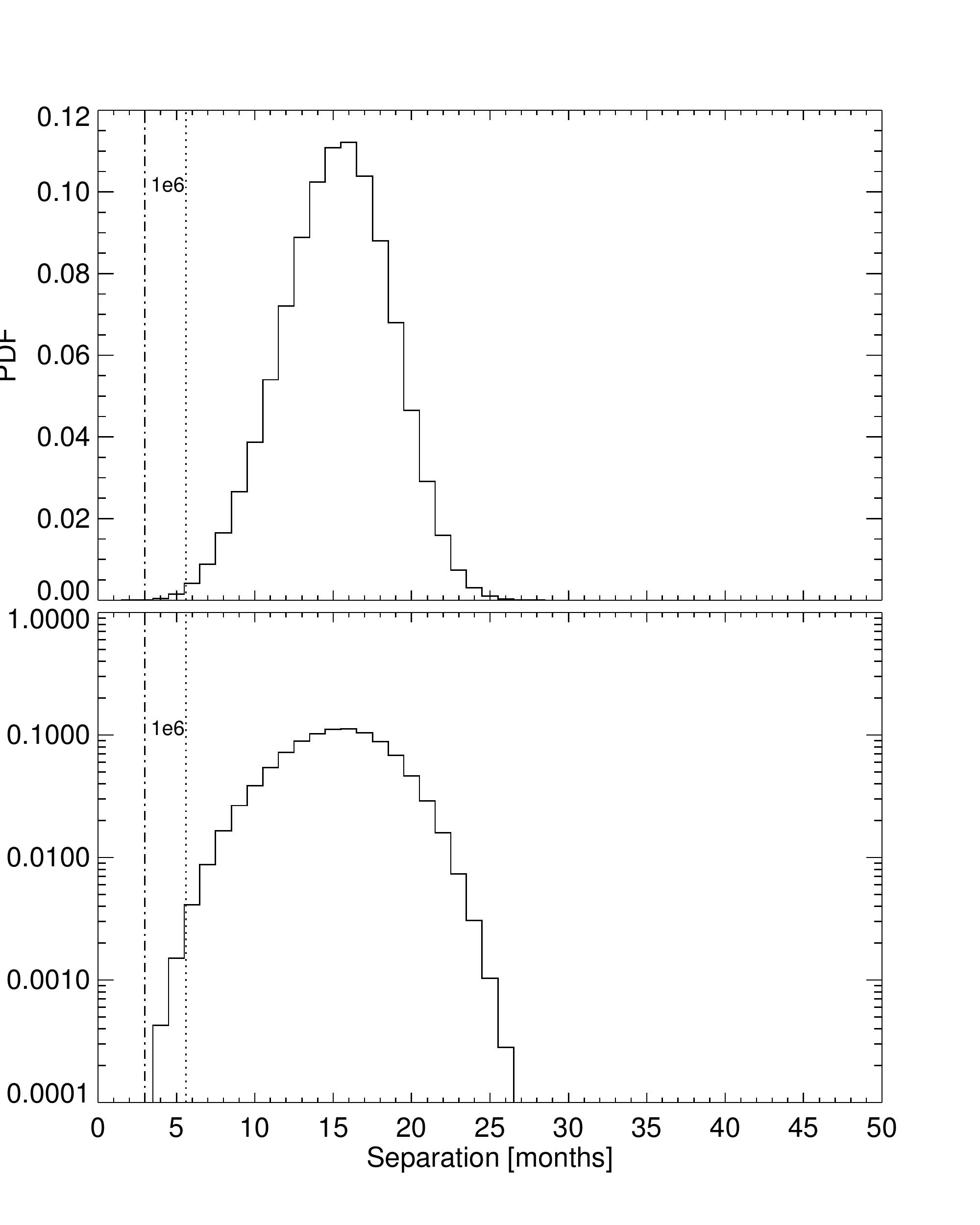}
\includegraphics[width=0.475\linewidth]{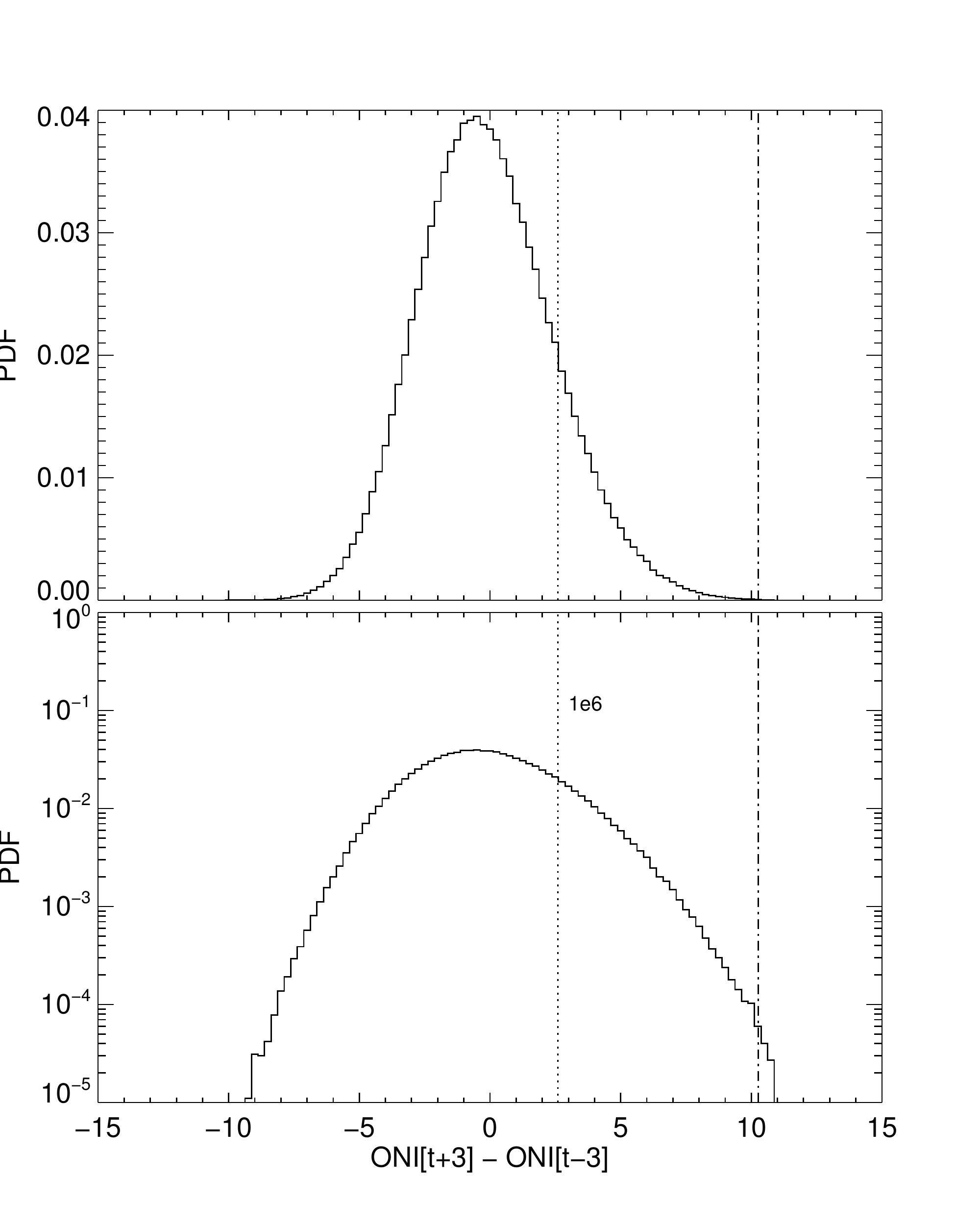}
\caption{
(left) Monte Carlo test 1: normally-distributed (in time) ENSO flips. 
The dotted line indicates 5.6 months and the dot-dashed line indicates 3.0 months---the observed and GCR-corrected mean terminator-ENSO flip separation (see text); smaller numbers are better.
(right) Monte Carlo test 2: piecewise-shuffled $\delta$ONI across terminators.
The dotted and dot-dashed lines again indicate the observed and GCR-corrected averages; in this case larger numbers are better.
In each case
the top panels show the simulation results with a linear $y$-axis;
the bottom panels show the same data on a logarithmic scale.
}
\label{fig:fMCT12}
\end{figure}

First, we observe from Fig.~\ref{fig:f3} that there are 13 major El Ni\~{n}o to La Ni\~{n}a transitions (defined as a change of the NOAA ONI index of $-1$ in less than 12~months) over the duration of the dataset;
the mean gap between them is $57.4 \pm 25.5$~months. Repeatedly creating an artificial 
``ONI'' 
time series akin to Fig.~\ref{fig:f3} with 13 El Ni\~{n}o to La Ni\~{n}a transitions, normally distributed, and computing the mean separation with the closest terminators, we find that in only 40 of $10^6$ runs is the mean terminator separation 3 months or less, and in only 1361 of $10^6$ runs is the mean terminator separation 5.6 months or less, 
where 5.6 months is the mean La Ni\~{n}a lag, uncorrected for Rossby-driven short-term fluctuations (see Table~\ref{tab:t1}).
By comparison, over all $10^6$ runs the mean terminator separation is $20.8\pm8.1$ months, 
as shown in the left-hand panels of Figure~\ref{fig:fMCT12}, where smaller separations are better.
Choosing a Poisson or uniform distribution instead of a normal distribution does not change the overall result, and in fact only decreases the number of runs where the mean terminator separation is less than observed reality---the Poisson distribution, probably the most realistic case, has only two out of $10^6$ runs better than reality.

The second Monte Carlo trial divides the observed ONI time series since 1960 into 32 pieces, with each positive-negative zero crossing of the time series defining those pieces. For each trial these pieces are randomly reordered, and the (summed) change in the ONI index from the observed Terminator dates $\pm 6$ months is compared to the observed ONI data. The results are shown in the right-hand panels of Figure~\ref{fig:fMCT12}:
In only 203 of $10^6$ runs is the simulated data better than the historical data. 

\begin{figure}[ht]
\centering
\includegraphics[width=\linewidth]{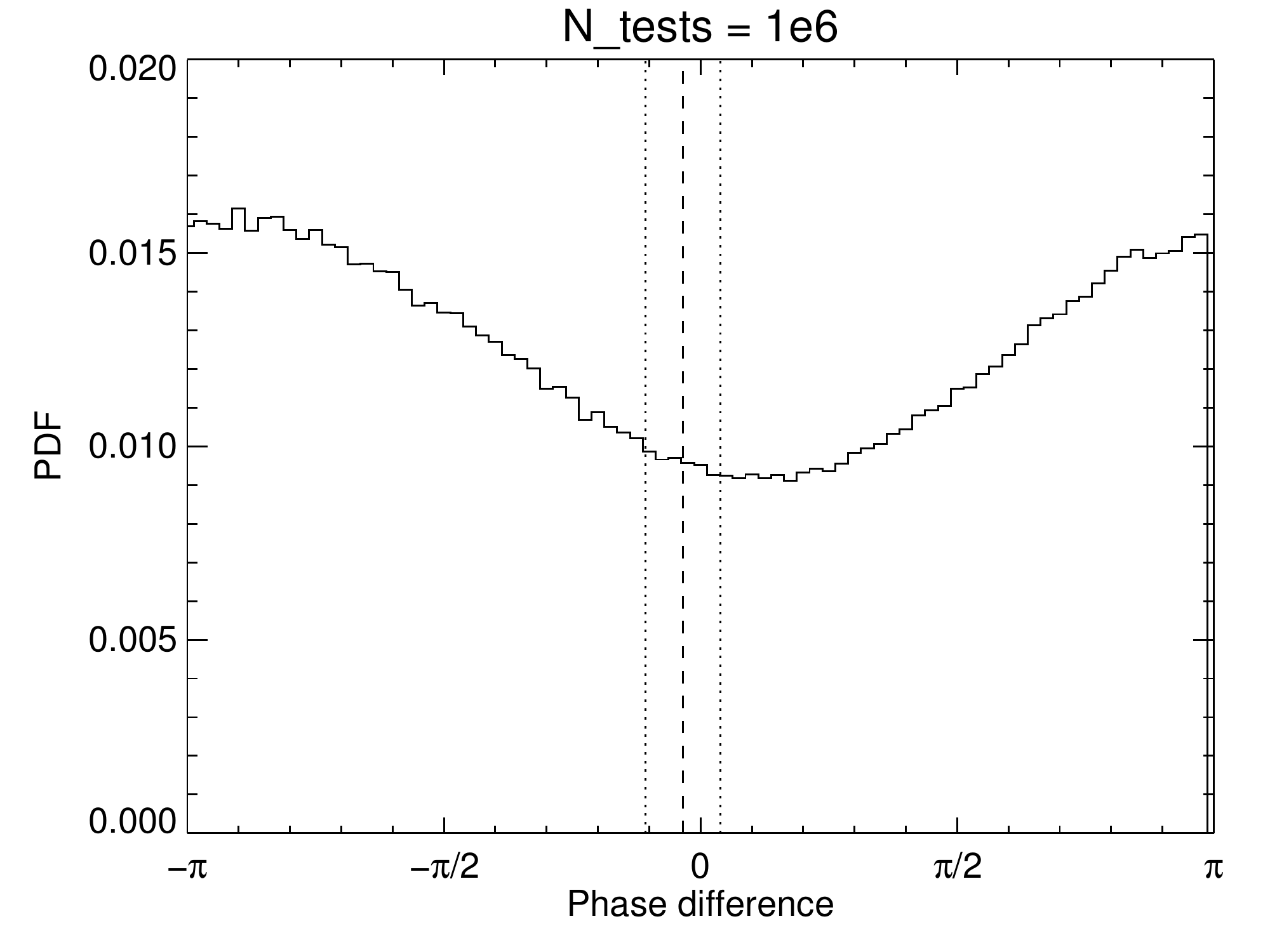}
\caption{Monte Carlo 3: piecewise shuffle Hilbert -- one-to-one correlated, if not causally connected.
The dashed line indicates the observed mean phase difference, and the dotted lines the standard deviation. Only 1265 of $10^6$ runs lie within the stdev envelope, which is consistent with zero.
}
\label{fig:fMCT3}
\end{figure}

{
The third Monte Carlo trial again takes the random reordering of 32 pieces of the ONI record, but instead of computing the summed change in the ONI index, 
which can be skewed by correctly guessing the largest ENSO flips (e.g., 1997--98),
we employ a Hilbert transform phase technique.
In signal processing, the Hilbert transform is a specific linear operator that takes a function, $u(t)$ of a real variable and produces another function of a real variable ${\cal H}[u(t)]$. 
This linear operator is given by convolution with the function 
$1/(\pi t)$.
The Hilbert transform has a particularly simple representation in the frequency domain: it imparts a phase shift of $90^\circ{}$ to every Fourier component of a function; 
as such, an alternative interpretation is that the Hilbert transform is a ``differential'' operator, proportional to the time derivative of $u(t)$. 
Thus a time series $z(t)$ can be expressed as $z(t) = u(t) + i {\cal H}[u(t)]$, where ${\cal H}[u(t)]$ represents the Hilbert transform of time series, $u(t)$. Equivalently, $z(t) = A(t) exp[i \phi(t)]$, where  $A(t)$ and $\phi(t)$ are the {\em instantaneous\/} amplitude and phase functions respectively of the time series \cite{Bracewell2000,2002AmJPh..70..655P,2018PhPl...25f2511C}.

It is that analytic temporal phase $\phi(t)$ that we refer to above as the Hilbert phase of ENSO variability.
It also follows from  
$\omega(t) = d\phi(t) / dt$ defining the slope of the changing phase with time has significance as a ``localized'' or ``instantaneous'' period of the fluctuating quantity.
A second useful feature of the Hilbert phase is in the phase coherence of two time series:
if edges/events in one time series occur at constant phase in another, the two are one-to-one correlated, or ``phase locked'' or ``synchronized'' \cite{2018PhPl...25f2511C,2018NucFu..58l6003C}.
\citeA{2019arXiv190906603L} showed that the Hilbert method could accurately and robustly determine the terminators from sunspot numbers alone.

This technique is used to calculate the mean phase difference between the  (shuffled) ONI crossings and the prescribed $N$ terminators, 
{\em i.e.,} $[\Delta\phi(t_0) + \Delta\phi(t_1) + \ldots + \Delta\phi(t_{n-1})]/N$, where $t_n$ is the time of the $n$th terminator.
The results of this method, for $10^6$ Monte Carlo runs are shown in Figure~\ref{fig:fMCT3}.
The dashed line is the mean phase difference of the 5 terminators (cycles 19--23) is $-0.035\pi \pm 0.073\pi$, and the dotted lines describe the uncertainty.

Thus, 
combining the results of these three different tests,
we can say with
with a confidence $p> 0.9986$, the correlation between terminators an ENSO crossings is {\em not\/} a coincidence.
}

\subsection{The ``Standard'' Cycle}
\label{sec:unit}

\begin{table}[t]
\centering
\begin{tabular}{|l|l|r|}
\hline
Cycle & Terminator Date &  Retro-prediction \\
{} & (Observed) & Skill Score (Percent) \\
{} & (Predicted) & {} \\
\hline
20 & 1978 Jan 01 &  30.3 \\
\hline
21 & 1988 June 01 & 29.1  \\
\hline
22 & 1997 Aug 01 & 53.6 \\
\hline
23 & 2009 Dec 15 & 54.0 \\
\hline
24 & 2020 Apr 01\ldots & 69.8$*$  \\
\hline
\end{tabular}
\caption{\label{tab:t2}
Skill Scores for using the ``standard'' ONI cycle of Fig.~\protect\ref{fig:fU} as prediction for each of the past 5 cycles. For the ongoing cycle 24, we only compute the prediction through 2018 June. 
}
\end{table}

\begin{figure}[ht]
\centering
\includegraphics[width=\linewidth]{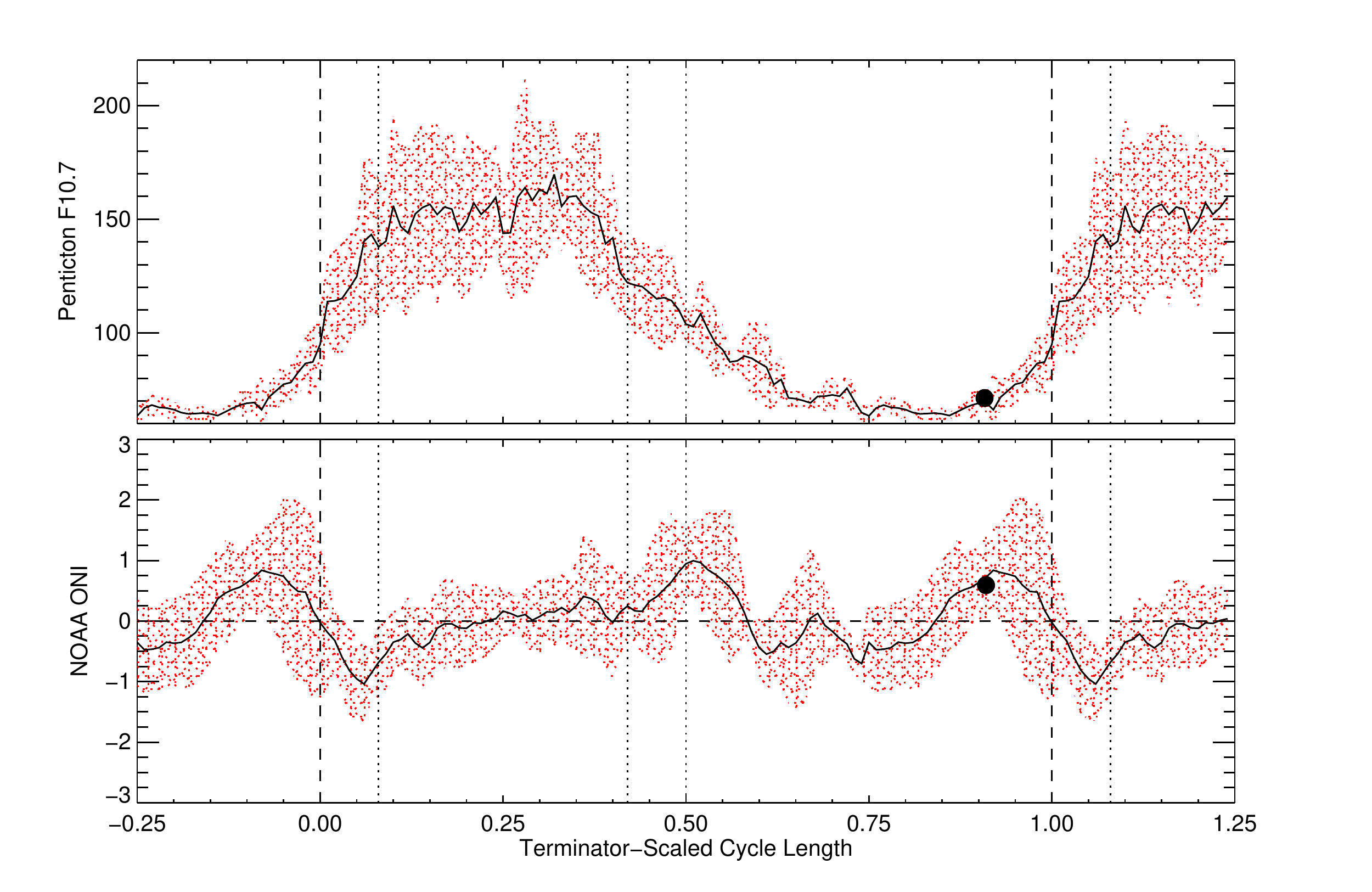} 
\caption{``Standard'' cycle for ONI (top) and F10.7 (bottom). 
The black trace is the average, and the red envelope is defined by one standard deviation.
The dots correspond to 2019 May, the latest measurements for Cycle 24 used in the computation, prior to submission.
}
\label{fig:fU}
\end{figure}

As previously discussed, 
it is clear from the modified superposed epoch analysis of Fig.~\ref{fig:fS2} that there is a coherent pattern to solar output and the terrestrial response from terminator to terminator.
The logical next step, then, is to average the five solar cycles for which we have data into a ``standard'' unit cycle that we may use for skillful prediction of future cycles.
The monthly series data are interpolated into 100 points from terminator to terminator, and an average and standard deviation are computed for each of 5 points for $x \leq 0.77$ and 4 points for $x > 0.77$; based on the terminator dates listed in Table~\ref{tab:t1}, June 2018 corresponds to $x = 0.77$. 
This is shown in Fig.~\ref{fig:fU} for F10.7 and the ONI El Ni\~{n}o index. 
Given the almost 100\%\ variation in peak F10.7 from cycle to cycle, the average rises more smoothly from solar minimum to solar maximum than any of the individual cycles of Fig.~\ref{fig:fS2}; 
however, the changes in standard deviation 
({\em i.e.}, the edges in the red shaded envelope) 
are clear at $x \sim 0$, 0.08, 0.42 and 0.50, and are driven by the Rossby waves in the solar tachocline as discussed in section~\ref{sec:flip} above.

We may use this standard cycle as a prediction tool for future ENSO events. As a test, we use the standard ONI cycle to retro-predict each of the past 5 cycles, and compute a skill score 
relative to a ``prediction'' of no oscillation (ONI constant at zero).
These hindcasts are shown in Table~\ref{tab:t2}. 
50\%\ is good and 70\%\ is an extremely good score, although with the caveat that we used these five cycles to compute the average; clearly Cycle 24 is closest to the standard average cycle.

\begin{figure}[ht]
\centering
\includegraphics[width=0.9\linewidth]{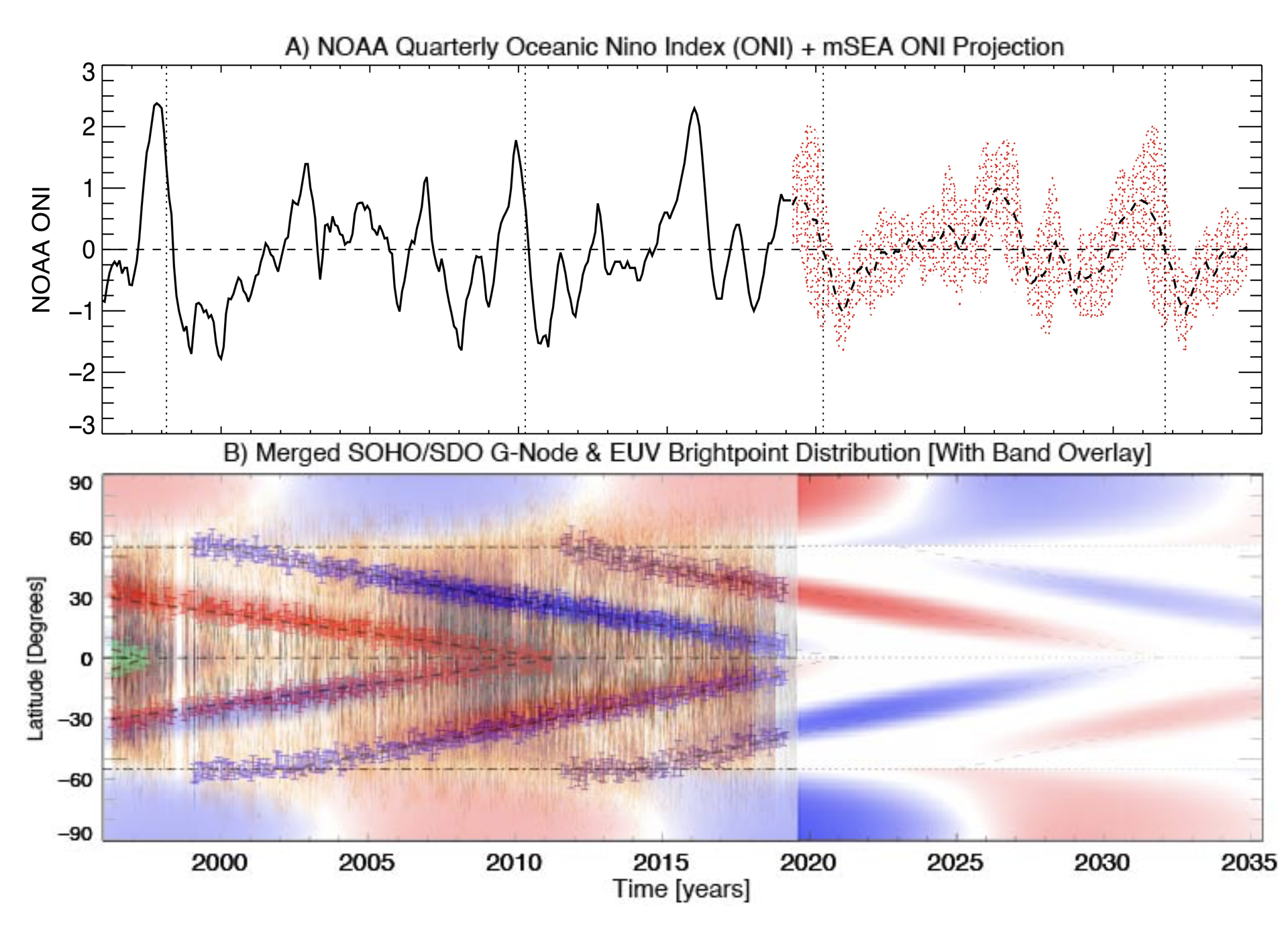}
\caption{``Standard'' cycle from Fig.~\protect\ref{fig:fU} projected forward in (real) time to the Cycle~25 terminator, currently predicted from Table~\protect\ref{tab:t3} be October 2031. El Ni\~{n}os may be expected around 2026 and 2031, and La Ni\~{n}as in 2020--21, 2027--28 and 2032--33. 
}
\label{fig:ffuture}
\end{figure}

\begin{table}[t]
\centering
\begin{tabular}{|l|rr|l|}
\hline
Cycle & Band Speed & (Degrees/ Year) & Terminator Date  \\
{} & (North) & (South) & (Observed/ Predicted)  \\
\hline
22 & $-2.17 \pm 0.17$ & $2.09 \pm 0.15$ & 1997 Aug 01 \\
\hline
23 & $-1.99 \pm 0.06$ & $2.00 \pm 0.04$ & 2009 Dec 15 \\
\hline
24 & $-3.05 \pm 0.11$ & $3.16 \pm 0.09$ & 2020 Apr 01\ldots  \\
\hline
25 & $-2.84 \pm 0.17$ & $2.84 \pm 0.20$ & 2031 Oct 01\ldots \\ 
\hline
\end{tabular}
\caption{\label{tab:t3}
Large scale equatorward motion of the activity bands through tracking the evolution of the brightpoint latitudinal density distributions. Cycles 24 and 25 fits calculated through September 1, 2018.}
\end{table}


In the language of the state vector simple dynamic system formulation of ENSO of \citeA{1995JCli....8.1999P}, it is clear that the forcing term $f(t)$ must have a strong negative impulse at the terminator, a (strong) positive impulse through sunspot minimum to the terminator (and one---or two, for each hemisphere---weaker positive impulses associated with increased (E)UV insolation around solar maximum).
As Fig.~\ref{fig:wavelet} shows
[and as \citeA{TorrenceCompo98} and \citeA{1996JCli....9.1586W} showed],
there is always power at shorter scales (3--7 years), between the terminators, corresponding to the intrinsic mode(s) of the system. 
Even if 
it is likely that 
the mid-cycle El Ni\~{n}o peak is related to increased solar irradiance,
as mentioned in Section~\ref{sec:flip}, we
do not attempt to fit every bump and wiggle, or explain every (non- terminator) feature as solar-induced.

Nevertheless, it is an interesting exercise, if not an acid test, to predict Cycle 25:
we {\em already\/} can estimate the date of the next terminator date as the brightpoints revealing the Cycle~25 activity band 
(cf.\ Fig.~\ref{fig:f1}b) 
have been present on disk long enough 
such that we may make a (well-constrained) linear extrapolation for when the Cycle~25 terminator will be
and thus convert the unit cycle to real time out beyond 2030.
This is shown in Fig.~\ref{fig:ffuture}:
The lower panel updates Fig.~\ref{fig:f1}b, and shows the progression of the EUV brightpoint distribution for cycles 22--25. 
That the cycle 25 progression is well-established and, more importantly, linear, is clear.
The linear fit of equatorward band progression
(Table~\ref{tab:t3}) is $2.84^\circ$/year, so the Cycle~25 terminator can be predicted to be in late 2031, with an uncertainty of $\pm 12$--15 months, given the present fit uncertainties.

If 
the Solar Cycle-CRF-ENSO connection described here holds for the Cycle~24 terminator in 2020, we may be cautiously optimistic for the general trends of large-scale global climate in the next decade.


%
\bibliographystyle{agufull08}

%

\end{document}